\documentclass{aa}  
\usepackage{natbib}
\usepackage{graphicx}

\usepackage{txfonts}
\usepackage[colorlinks=true,linkcolor=blue,citecolor=blue]{hyperref}
\usepackage{siunitx}
\usepackage{appendix}

\usepackage{placeins}

\begin{document}

   \title{Photometric classification of stars around the Milky Way's central black hole}

   \subtitle{I. Central parsec}

   \author{E. Gallego-Cano
          \inst{1}
         \and
           T. Fritz
          \inst{2}
          \and
          R. Sch\"odel
          \inst{1}
          \and
          A. Feldmeier-Krause
          \inst{3}
          \and
          T. Do
          \inst{4}
          \and
          S. Nishiyama
          \inst{5}
   }

   \institute{
       Instituto de Astrof\'isica de Andaluc\'ia (CSIC),
     Glorieta de la Astronom\'ia s/n, 18008 Granada, Spain; \email{lgc@iaa.es}
     \and
     Department of Astronomy, University of Virginia, 530 McCormick Road, Charlottesville, VA 22904, USA
        \and
        Department of Astrophysics, University of Vienna, T\"urkenschanzstraße 17, Wien, 1180, Austria
        \and
        Physics and Astronomy Department, University of California, Los Angeles, CA 90095-1547, USA
         \and
        Miyagi University of Education, Sendai, Miyagi, Japan
             }

   \date{35 pages, 38 figures, accepted for publication in A\&A}

 
  \abstract
   {The presence of young massive stars in the Galactic Centre (GC) raises questions about how such stars could form near the massive black hole Sagittarius\,A* (Sgr\,A*). Furthermore, the shape of the initial mass function (IMF) in this region seems to differ from its standard Salpeter/Kroupa law. Due to observational challenges such as extreme extinction and crowding, our understanding of the stellar population in this region remains limited, with spectroscopic data available only for selected small and comparably bright sources.}
   {We aim to improve our knowledge about the distribution and the IMF of young, massive, stars in the vicinity of Sgr\,A*.}
   {We used intermediate band (IB) photometry to identify candidates for massive young stars. To ensure robust classification, we applied three different, but complementary methods: Bayesian inference, a basic neural network, and a fast gradient-boosted trees algorithm.}
   {We obtain spectral energy distributions for 6590 stars, 1181 of which have been previously classified spectroscopically. We identify 351 stars that are classified as early types by all three classification methods, with 155 of them being newly identified candidates. The radial density profiles for late and early-type stars fit well with broken power laws, revealing a break radius of $9.2\pm 0.6''$ for early-type stars. The late-type stars show a core-like distribution around Sgr\,A* while the density of the early-type stars increases steeply towards the black hole, consistent with previous work. We infer a top-heavy IMF of the young stars near Sgr\,A* ($R < 9''$), with a power-law of $1.6\pm 0.1$. At greater distances from Sgr\,A* a standard Salpeter/Kroupa IMF can explain the data. Additionally, we demonstrate that IB photometry can also constrain the metallicities of late-type stars, estimating metallicities for over 600 late-type stars. 
   }
   {The variation of the IMF with radial distance from Sgr\,A* suggests that different mechanisms of star formation may have been at work in this region. The top-heavy IMF in the innermost region is consistent with star formation in a disc around Sgr\,A*.}

   \keywords{Galaxy: center --
               Galaxy: kinematics and dynamics --
                Galaxy: nucleus
               }

   \maketitle
%

\section{Introduction}\label{sec:intro}

As the nearest galaxy nucleus, situated at only ${\sim}8$\,kpc from Earth \citep{do2019relativistic, abuter2019geometric,abuter2021improved}, the Galactic Centre (GC) harbours in its centre a supermassive black hole Sagittarius\,A* (Sgr\,A*) with $4.04\pm 0.06\times10^{6}$\,M$_{\odot}$ \citep{abuter2018detection, do2019relativistic} surrounded by a nuclear star cluster (NSC) of $2.5 \times 10^{7}$\,M$_{\odot}$ \citep{abuter2018detection, do2019relativistic, fritz2016mass, feldmeier2017triaxial,schodel2014surface,launhardt2002nuclear} and a half-light radius or effective radius of ${\sim}4.2-7$ pc \citep{schodel2014surface,fritz2016mass,alexander2017stellar,gallego2020new}. Thanks to their nearness, the stars can be resolved observationally on scales of milliparsec inside the radius of influence of the black hole \citep{alexander2005stellar, gallego2018distribution, Schodel:2018db}. The Centre of the Milky Way therefore provides a unique possibility to study the stellar population in the most extreme astrophysical environment of our Galaxy and to study the interaction of stars with a massive black hole. 

Over the last few decades, numerous studies have unveiled the presence of young (${\sim}2-8$ Myr old), massive stars within ${\sim}0.5$\,pc of the black hole \citep[e.g.][]{ghez2003first, paumard2006two, lu2008disk,bartko09warped,genzel2010galactic, do2013densities, yelda2014clockwise}. This discovery is unexpected, as star formation is typically thought to be suppressed in the vicinity of a massive black hole due to its strong tidal field. There is a peculiar population of apparently main-sequence B stars within 0.08$''$ of the black hole, called `S-cluster'. At a larger radius, the young stars are distributed in two main features: the clockwise (CW) rotating disk in the region ${\sim}1$$''$ (${\sim}0.04$pc) to ${\sim}8$$''$ (${\sim}0.32$) from the black hole, and the counterclockwise (CCW) disk at distances larger than $8''$ (${\sim}0.32$pc) from Sgr\,A*. However, the latter feature is controversial, with some studies supporting its existence \citep{genzel2003stellar, paumard2006two, bartko09warped} while others refute it \citep{lu2008disk, yelda2014clockwise}. While all studies find a top-heavy mass function for the young stars near Sgr\,A*, there are significant discrepancies as to its exact power-law slope, from a value of -0.45 reported by \cite{bartko10imf} to -1.7 reported by \cite{lu2013stellar}. This inconsistency may arise from the limited and different completeness of the data in these studies. Recently, \citet{fellenberg2022young} examined an extensive area (${ \sim }30'' \times 30''$), the most extensive study of young stars in the GC to this date. They confirm the presence of the warped CW disk, an outer kinematic feature (F2), and the CCW feature at larger distances. Additionally, a previously unreported feature (F3) was identified beyond a projected radius of $R=8''$ from Sgr\,A*. The stars from the warped CW disk exhibit a top-heavy mass function, in contrast to the stars at larger distances.

An important question regards the formation mechanism of the young stars. While most studies support in situ star formation from a massive gaseous disk, none can provide a complete explanation. Identifying more early-type stars will enhance our comprehension of their distribution and, consequently, shed light on their initial mass function (IMF), potentially providing further insights.

Given the extreme interstellar extinction and the extreme source crowding towards the GC \cite[e.g.][]{nishiyama2009interstellar, schodel2010peering,fritz2011line, nogueras2018galacticnucleus} it is extremely challenging to study the stellar population near Sgr\,A*. The classification of stars is primarily carried out using spectroscopy with Adaptive Optics (AO)-assisted high-angular resolution Integral Field Units, such as SINFONI/ERIS at the ESO-VLT or the OSIRIS spectrograph at the W.\ M.\ Keck observatory. In the most recent study by \cite{fellenberg2022young}, the analysis is limited to stars brighter than 15 magnitudes in the $K_{S}$-band.  

Several studies suggest that conducting imaging observations using intermediate-band (IB) photometry in the $K_{S}$-band \cite[e.g.][]{nishiyama2013young, nishiyama2023search,buchholz2009composition,plewa2018random} can be a powerful approach for identifying young and intermediate-age stars in the GC. These studies offer a significant advantage in detecting fainter stars within larger fields compared to spectroscopic studies. Nonetheless, it is crucial to emphasize that subsequent in-depth spectroscopic follow-up observations are essential for confirmation.

In this work, we use seven intermediate bands covering the near infrared (NIR) obtained by the AO-assisted NIR camera NACO installed at the ESO-VLT. We use the same data as \cite{buchholz2009composition} but apply an improved methodology and analysis. In recent years, the data has also been reanalyzed by \cite{plewa2018random}, who used a machine-trained classifier.

Our work aims to improve on and go beyond the previous work in several aspects:

\begin{enumerate}
    \item We study a somewhat larger region, covering an area over 12\% larger than that examined in previous studies, due to the exclusion of the H-band from our analysis, which has a smaller field of view (FoV).
    \item We enhance the image reduction process.
     The incorporation of noise maps and the utilization of a spatially variable PSF, not only results in deeper data but also improves the quality of the photometry.
    \item We determine robust photometric uncertainties by applying a bootstrapping procedure and including PSF uncertainties.
    \item We explore the effect of different models to fit the measured spectral energy distributions (SEDs).
    \item We employ three distinct yet complementary classification methods, encompassing Bayesian inference, a basic neural network, and a fast gradient-boosted trees algorithm. 
    \item Thanks to numerous spectroscopic studies performed in the past decade, we can use hundreds of classified stars over a large field, which leads to increased precision and accuracy of our classification.
\end{enumerate}

Our approach allows us to identify new candidates for early stars near Sgr\,A*. With these new data, we study the surface-density profile, offering a glimpse into the dynamical state of the cluster, and the luminosity function, a fundamental parameter that can be employed to determine properties such as age, star formation history, and the IMF of the cluster. In Section \ref{sec:data_reduction}, we outline the data reduction process and highlight the novel improvements made. Our photometric analysis, along with astrometric and photometric calibration, is detailed in Section \ref{sec:data_analysis}. Section \ref{sec:bayesian} presents our application of Bayesian inference for stellar classification. Additionally, two machine-learning methods—Multi-Layer Perceptron and XGBoost—are discussed in Section \ref{sec:ML}. We summarize the results in Section \ref{sec:result} and deliberate them further in Section \ref{sec:discussion}. We present our conclusions in Section \ref{sec:conclusions}. In this paper, we adopt a distance to Sgr\,A* of 8.28 kpc \citep{abuter2021improved}, with $1''$ corresponding to approximately 0.04 pc.

\section{Data reduction}\label{sec:data_reduction}
\subsection{Basic reduction} \label{sec:basic_reduction}
The data were obtained by using seven IB filters in $K_{S}$-band obtained with the S27 camera of NACO/VLT, with a pixel scale of $0.027''$. The AO was locked on the NIR bright supergiant GCIRS\,7 that is located about $5.5''$ north of Sgr\,A*. The data used are summarised in Table\,\ref{Tab:Obs}.  

Most of the data were acquired with a similar rectangular dither pattern, roughly centred on Sgr\,A*, but some of them were subjected to random dithering. The observations are described in detail in \cite{buchholz2009composition}. In contrast to \cite{buchholz2009composition}, we chose not to include H-band data in our analysis because the area observed by the H-band data has a more limited FoV in comparison to the IB data. Furthermore, as discussed in Section \ref{sec:discussion}, this data is not essential for our analysis.

We applied standard data reduction, with sky subtraction, bad pixel removal, and flat fielding. Afterwards, we aligned the individual images with the $K_{S}$-band wide-field mosaic from 11th May 2011 obtained with the S27 camera of NACO and described in \cite{gallego2018distribution}. The final mosaics for each epoch were created by mean-combining the individual exposures. The corresponding noise maps were computed using the error of the mean. Before mosaicing, we had rebinned the data with a factor of two and quadratic interpolation, which can improve the astrometry and photometry of the final product \citep[see][]{gallego2018distribution, Schodel:2018db}. We present the KLF for the final mosaic images and the photometric uncertainties in the Appendix \ref{unc_estimation}. The quality of the different IB data varies: IB224, IB233, IB227, and IB230 have a crowding limit of around 19 mag., IB200 and IB236 around 17.8 mag., and IB206 around 19.5 mag.

\begin{table}
\centering
\caption{Details of the imaging observations used in this
  work.}
\label{Tab:Obs} 
\begin{tabular}{llllll}
\hline
\hline
Date$^{\mathrm{(a)}}$ & $\lambda_{\rm central}$ & $\Delta\lambda$ & N$^{\mathrm{(b)}}$ & NDIT$^{\mathrm{(c)}}$ & DIT$^{\mathrm{(d)}}$\\
 &  [$\mu$m]  &   [$\mu$m] &  & & [s] \\
\hline
09 July 2004 & 2.00 & 0.06 & 8 & 4 & 36  \\
12 June 2004 & 2.06 & 0.06 & 96 & 1 & 30  \\
12 June 2004 & 2.24 & 0.06 & 99 & 1 & 30  \\
09 July 2004 & 2.27 & 0.06 & 8 & 4 & 36  \\
09 July 2004 & 2.30 & 0.06 & 8 & 4 & 36  \\
12 June 2004 & 2.33 & 0.06 & 119 & 1 & 30  \\
09 July 2004 & 2.36 & 0.06 & 8 & 4 & 36  \\
\hline
\end{tabular}
\tablefoot{
$^{\mathrm{(a)}}$ UTC date of beginning of night. $^{\mathrm{(b)}}$ Number of (dithered) exposures $^{\mathrm{(c)}}$ Number of integrations that were averaged on-line by the read-out electronics. $^{\mathrm{(d)}}$ Detector integration time. The total integration time of each observation amounts to N$\times$NDIT$\times$DIT. The approximate spatial resolution of the images is 0.07$''$-0.10$''$/pixel. We refer to the different IB bands by adding the number of their central wavelength: IB200, IB206, IB224, IB230, IB233, and IB236.}

 \end{table}

 \subsection{Repairing saturated stars} \label{sec:saturated_stars}
Repairing saturated stars is essential for obtaining accurate photometric measurements because the brightest stars will provide the most accurate estimates of the point spread function (PSF) wings. We repaired the saturated cores of bright stars with the PSF of nearby unsaturated stars, a methodology based on the {\it StarFinder} IDL routine REPAIR\_SATURATED \citep{diolaiti2000analysis}. The main difference to the latter is that we also applied an additive offset when matching the PSF from unsaturated stars to one of the saturated stars, which is necessary because of the different normalization factors of the PSFs. The selection of the PSF reference stars is described in Section \ref{sec:psf_fitting}.

\section{Data analysis}\label{sec:data_analysis}

\subsection{PSF fitting} \label{sec:psf_fitting}

Stellar photometry and astrometry were acquired with the PSF fitting program
{\it StarFinder} \citep{diolaiti2000analysis}. The FoV of all images is $28''\times28''$, larger than the isoplanatic angle in the $K_{S}$-band, therefore special attention was given to addressing the spatial variability of the PSF by using local PSFs. We divided the images into sub-fields of approximately $10.8''\times10.8''$ size following the procedure described in \cite{gallego2018distribution}. Since anisoplanatic effects have a relatively limited impact on the seeing halo, as noted \citet{schodel2010accurate}, we employed the brightest star, $GCIRS7$, to estimate the seeing halo for the PSF of all sub-fields \cite[see] []{gallego2018distribution}.

We first removed any extended sources from the list of detected stars and then selected the ten brightest and most isolated stars to perform an iterative extraction of the local PSF within each sub-field.
Differences in mean extinction and source density between the sub-fields imply that not all of them contain reference stars with uniform brightness, leading to small systematic offsets in the zero-points. We address this problem in section\,\ref{sec:local_calibration}. The positions and fluxes of each star were determined by averaging multiple measurements from overlapping frames.

Finally, we applied the bootstrapping resampling method to compute the photometric uncertainties. This method allowed us to obtain robust and reliable uncertainties in a crowded field without loss of sensitivity, as described by \cite{gallego2022accurate}. The details of the process are included in Appendix \ref{unc_estimation}.

To obtain the final list of stars, we combined the final lists of the seven IB-band data sets into a single list containing stars that were detected in every filter. The quality of the data is very different for the different bands, as we see in Fig.\, \ref{Fig:klf_normalboots}, and \ref{Fig:klf_noiseboots}. The number of stars that were found in all filters is limited to $6590$ because interstellar extinction rises steeply towards shorter wavelengths. 

\subsection{Astrometric calibration} \label{sec:astro_calibration}

In order to obtain the astrometric solutions, we used the radio positions from eight maser stars ($IRS9$, $IRS12N$, $IRS28$, $SiO\text{-}15$, $IRS10EE$, $IRS15NE$, $IRS17$, $IRS19NW$) in the FoV \citep{Reid:2007vn}. We applied a linear solution to obtain the World Coordinate Systems (WCSs) and to convert the pixel position to celestial coordinates, right ascension, and declination (IDL Astrolib routine SOLVE\_ASTRO).

\subsection{Photometric calibration} \label{sec:photo_calibration}
In this section, we describe the two steps involved in our photometric calibration. First, we applied a basic calibration to convert star fluxes into magnitudes. Second, we applied a local calibration to take into account zero-point variations across the field. While our approach is similar to that followed in \cite{buchholz2009composition}, we have included some essential differences to improve the procedure.

\subsubsection{Basic calibration} \label{sec:basic_calibration}
\begin{figure} [ht!]
\includegraphics[width=\columnwidth]{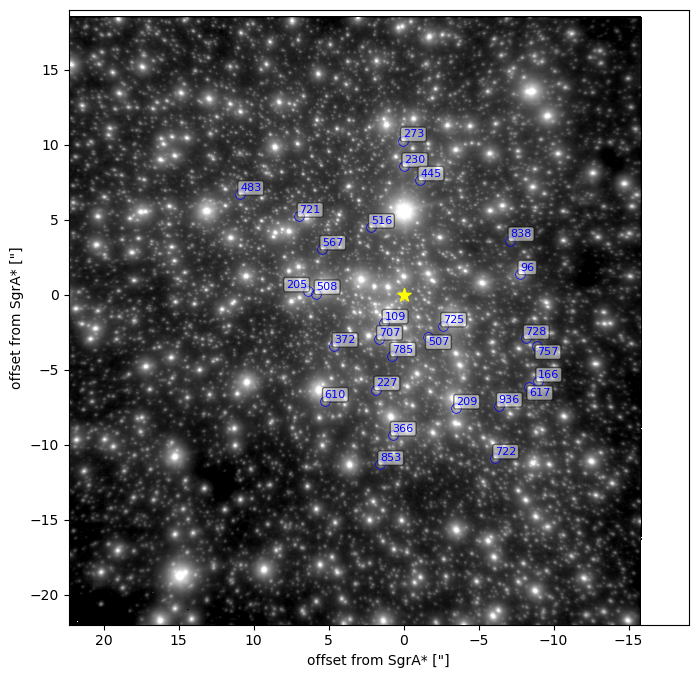}
\caption{Calibration stars used for the basic calibration. The labels indicate the identification numbers corresponding to the list of \cite{feldmeier2015kmos}. The axes show the offset distances from Sgr\,A* (yellow star) in arcsecond. The background image is from the 2004 June 12 observation corresponding to the IB224 filter. The FoV of the image is ${\sim}42''$ x $42''$(${\sim}$1.7pc x 1.7pc).}
  \label{Fig:map_calibrators}
\end{figure}

We chose OB stars as photometric calibrators because they can be closely approximated as blackbodies and lack distinctive features in their SED at the wavelengths of our data  \citep[see Fig.\,4 in][]{feldmeier2015kmos}. We selected OB stars from \cite{feldmeier2015kmos}, and removed the variable stars identified by \cite{gautam2019adaptive}. Subsequently, we selected isolated and bright stars, resulting in 28 OB calibrators, shown in  Fig.\,\ref{Fig:map_calibrators} and listed in Table\,\ref{Tab:calibrators}. 

We computed an extinguished blackbody SED with an effective temperature $T_{\text{eff}}$ of ${\sim}\SI{30000}{\K}$ for each OB star, taking its $A_{K}$ extinction value from \cite{schodel2010peering}. The absolute mean extinction value in $K_{S}$ is around 2.54. We assumed that the extinction curve in the NIR can be approximated by a power law, as suggested by previous studies \citep{nishiyama2009interstellar,fritz2011line}. Specifically, we chose $\alpha = 2.21$ \citep{schodel2010peering,nogueras2019variability}. Subsequently, we multiplied the blackbody SEDs with the transmission curves of the IB filters and integrated them in each band. These theoretical fluxes were then utilized for the comparison with the observed fluxes, from which we derived seven IB zero points for each OB star. The 28 zero points for each IB band were median combined to obtain the best zero point estimates.

In order to fix the absolute magnitudes, we assigned the $K_{S}$-band magnitudes from \cite{schodel2010peering} to the IB224 magnitude because it is the most central magnitude among all IB filters and remains unaffected by emission or absorption features. The selection of the same $T_{\text{eff}}$ for all reference stars does not affect the final calibration due to the excellent approximation provided by the Rayleigh-Jeans law for the SED of hot stars in the near-infrared region. We tested different values for $T_{\text{eff}}=15000,$ $35000$\,K, and the variation in calibrated magnitudes was $\leq 0.01$, $0.002$, respectively.

\begin{figure*}[ht]
\begin{center}
\includegraphics[width=0.66\columnwidth]{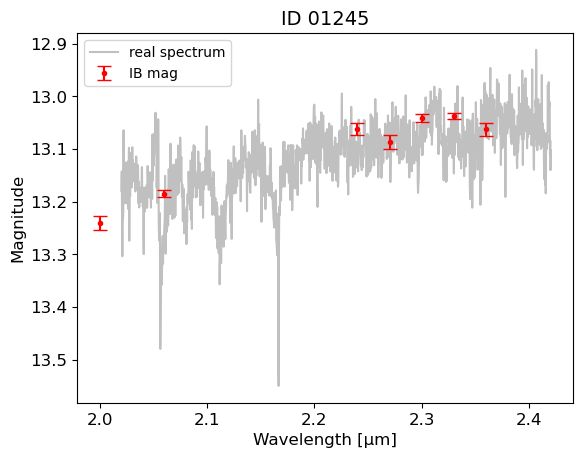}
\hspace{0.03cm}
\includegraphics[width=0.66\columnwidth]{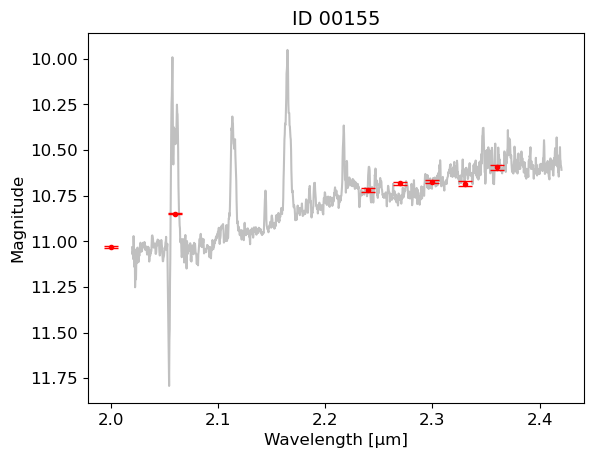}
\hspace{0.03cm}
\includegraphics[width=0.66\columnwidth]{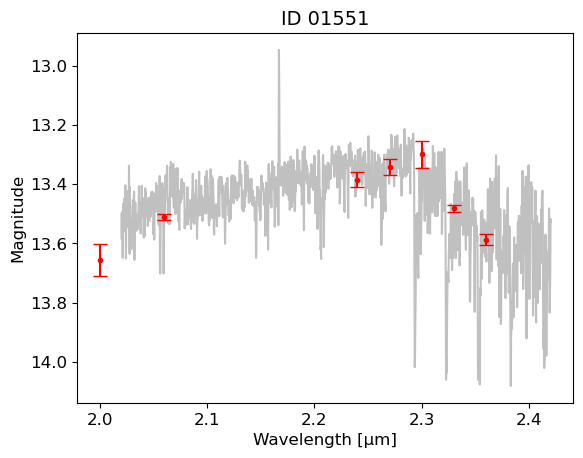}
\end{center}
\caption{Comparison between real spectra in $K$-band and the low-resolution SEDs computed with IB photometry (red points, optimally shifted to match the spectra ). The identification numbers are indicated in the panels. The spectra displayed in the left and middle panels correspond to an OB, and a Wolf-Rayet star, respectively, from \cite{feldmeier2015kmos}. The right panel shows a spectrum for a late-type star from \cite{feldmeier2017kmos}.}
  \label{Fig:real_spectra}
\end{figure*}

To assess the quality of the calibration, we compared the low-resolution SEDs computed with IB photometry with real spectra in the $K$-band of known stars. Figure\,\ref{Fig:real_spectra} shows the spectra of real stars situated in the innermost region, and their corresponding IB magnitudes (red points). The left and middle panels correspond to spectra from individual stars from \cite{feldmeier2015kmos} that are identified as early-type OB, and Wolf-Rayet stars, respectively. The right panel corresponds to a spectrum of a late-type star from \cite{feldmeier2017kmos}. The identification numbers are indicated in the panels. The SEDs follow very well the main features of the spectra. We can see that the main characteristic for distinguishing between early-type and late-type stars with our low resolution in these wavelengths is the CO bandhead feature from $K_{S} \geq 2.3\mu m$. In Section\,\ref{sec:cbd}, we explore various approaches to measure the depth of the feature for use in our analysis when classifying the stars.

\subsubsection{Local calibration} \label{sec:local_calibration}

We analysed the photometric calibration quality across the FoV because factors such as the distance to the AO guide star, AO performance, and the number of exposures per pixel are functions of the position in the field. In particular, the inner regions are deeper than the outer regions in the final mosaics because of a larger number of observed frames.

We divided the FoV into 25 squares of around $8.6''$ x $8.6''$ size (see Fig.\,\ref{Fig:sed_rc_fov}). We selected stars with magnitudes between the $K_{S}$ luminosity function (KLF) peak and $\pm 0.5$ magnitudes for each square, which, in most cases, are Red Clump (RC) stars. We investigated the mean SED of the RC stars within the sub-fields in Appendix\,\ref{SED_fov}, and discovered variations in the expected SEDs for these types of stars, especially in the outer regions, as depicted in Fig.\, \ref{Fig:sed_rc_fov} (blue points). Therefore, we applied a local calibration with dual purposes: firstly, to address systematic errors associated with the distance to the calibrator stars and image quality, as previously noted, and secondly, to correct for differential extinction. This secondary calibration is different from the approach of  \cite{buchholz2009composition}, but similar to the method used by \citet{plewa2018random}.

\begin{figure} [ht!]
\centering
\includegraphics[width=0.9\columnwidth]{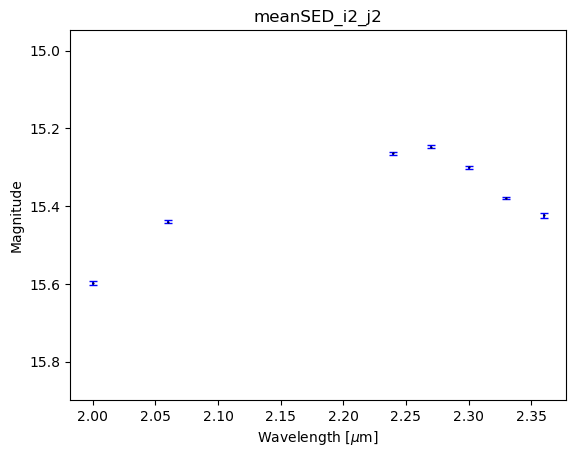}
\caption{Mean SED for $29$ RC stars selected in the inner region used as the template for local calibration. The considered stars are classified late-type by \cite{gautam2019adaptive}. Error bars indicating the standard errors are also depicted.}
  \label{Fig:template_SED}
\end{figure}

The calibration within the central region (i2, j2 in Fig.\,\ref{Fig:sed_rc_fov}) exhibits minimal sensitivity to systematic errors and statistical uncertainties because of the large number of exposures, closeness to the AO guide star and high density of calibration stars. We selected $29$ RC stars classified by \cite{gautam2019adaptive} in this region and derived their mean SED, which was subsequently employed as our SED template for the next step (see Fig.\,\ref{Fig:template_SED}).

For the local calibration, we selected RC stars across the FoV by considering the stars with magnitudes in the IB224 band between the peak of the KLF (around $15.6$ mag) and $\pm 0.5$ magnitudes. We excluded stars with significant photometric errors exceeding $0.5$ magnitudes. The total number of selected RC stars is $2742$. For each star in our list, we then created a mean SED from the 20 closest RC stars and determined the correction for each filter from the differences between the local RC SED and the template created from the RC stars in the central region. Finally, the magnitude corrections were applied to each of the corresponding IB magnitudes of the star. This local calibration procedure also corrects for differential extinction. In Fig.\,\ref{Fig:sed_rc_fov}, we compare the mean SED of RC stars in the sub-fields along the FoV. The blue points represent the SED after the basic calibration, while the red points depict the SED after applying the local calibration. The former is shifted to achieve overlap. The local calibration process enables us to obtain the final list of stars corrected for differential extinction while homogenizing the calibration quality across the entire FoV for different bands.

\subsection{CO band} \label{sec:cbd}

Late-type stars display a characteristic CO band head absorption at wavelengths $\lambda>2.27$ $\mu$m (see the right panel in Fig.\,\ref{Fig:real_spectra}). By examining the observed and calibrated SEDs for this feature, we can categorize stars as late-type candidates (spectral types $\sim$GKM) when it is observed, or as early-type candidates (spectral types $\sim$OB) when it is absent. The first four IB filters ($\lambda=2.00-2.27$ $\mu$m) are used to estimate the continuum.

To measure the CO band depth (CBD), we employed two different SED fitting methods. The first method was similar to \cite{buchholz2009composition}, but we fitted a straight line to the first four data points ($\lambda=2.00-2.27$ $\mu$m) instead of an extinguished black body, as they did. We also tested the latter method but found a straight-line fit to be more robust and to result in more reliable uncertainties, in particular for faint stars. Subsequently, we applied a third-order polynomial to the entire SED, replacing the first four data points with the fitted straight line to ensure a consistent fit. 

In the second method, we similarly fitted a straight line to the first four data points, followed by an exponential function represented by the equation:
\begin{equation}
\label{eqn:exp_fit}
f(x)=a+b\cdot x+10^{-0.4\cdot c}\cdot e^{x\cdot d},
\end{equation}

where $a$, $b$, $c$, and $d$ are the model parameters. We used the exponential function in two steps. First, we applied an exponential fit to the stars. Then, in the second step, we performed another fit, setting the $d$ parameter to the median value of $10.33$ obtained in the first step for classified late-type stars. This was done to prevent convergence issues, especially for stars with significant uncertainties.

\begin{figure*}[!ht]
\centering
\includegraphics[width=\textwidth]{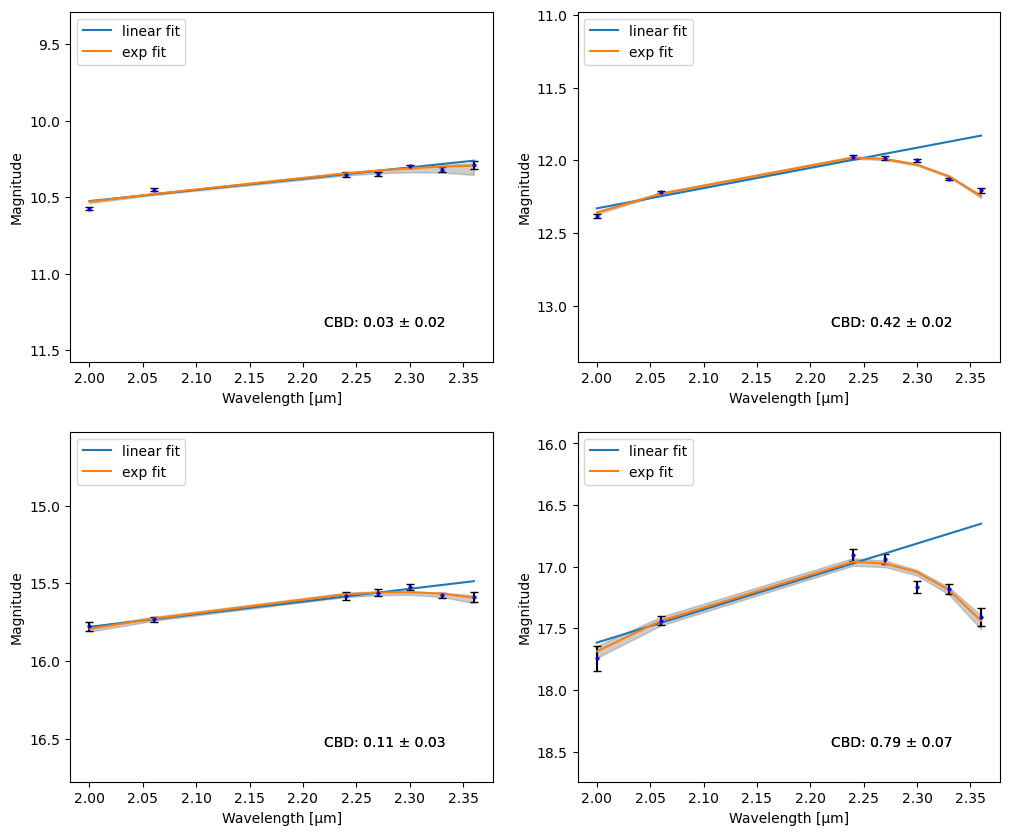}
\caption{SEDs of four spectroscopically classified stars. The top panel displays bright stars, and the bottom panel shows faint stars, with early-type stars on the left and late-type stars on the right. The blue lines denote the linear fit to the first four data points, while the orange lines represent the exponential fit to all seven data points. The grey shadows indicate the uncertainties of the exponential fits, as obtained from Monte Carlo simulations. The panels include notations for the CBDs for the stars.}
  \label{Fig:comp_CBD}
\end{figure*}

Finally, we computed the CBD values for both methods by subtracting the extrapolated values of the linear fit at $\lambda=2.36$ $\mu$m from the extrapolated values of the exponential fit, and the polynomial fit, respectively, at the same wavelength. Figure\,\ref{Fig:comp_CBD} shows the SEDs and their corresponding fits by using the exponential fit for four stars classified spectroscopically as early (left panel) and late (right panel), respectively. The figure includes the fits for bright stars (top panel), and for faint stars (bottom panel). The CBD values are indicated in the panels.

We employed a Python-based\footnote{Python Software Foundation. Python Language Reference, version 3.9. Available at \url{http://www.python.org}} procedure using the \texttt{curve\_fit} tool provided by the \texttt{SciPy}\footnote{\url{https://scipy.org/}} package, utilizing non-linear least squares to obtain optimal parameter values in both cases and including the photometric uncertainties in the fit. In Appendix\, \ref{CBD_fits_analysis}, we compare both methods, demonstrating that while the fits are very similar for most stars, significant differences emerge in cases where the SED exhibits notable irregularities (see Appendix \ref{CBD_fits_analysis} for detailed insights). The results obtained using exponential fits are more consistent, leading us to consider the CBD values derived from this method.

\begin{figure} [ht!]
\includegraphics[width=\columnwidth]{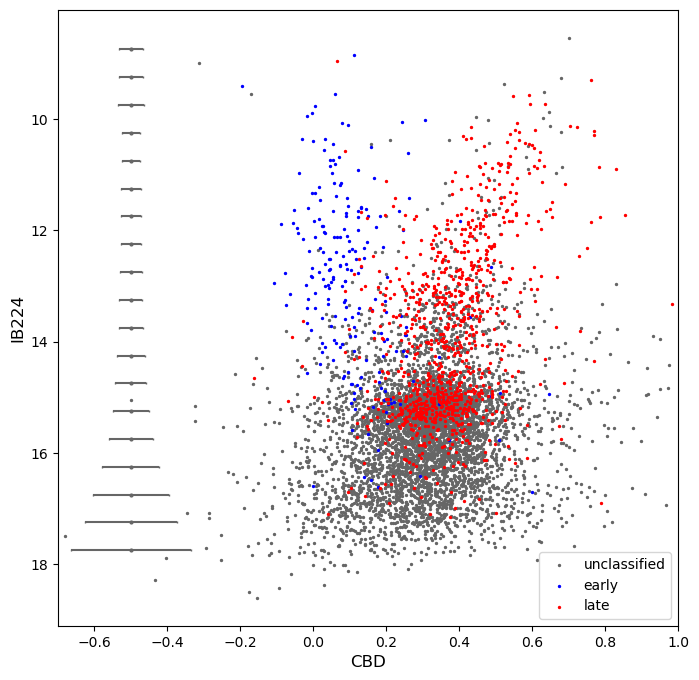}
\caption{CBD diagram for the entire sample. The median errors of the CBD for bins of 0.5 magnitudes are represented by grey bars on the left side of the panel. The spectroscopically classified stars are marked by blue (early-type) and red (late-type) points.}
  \label{Fig:final_cbd}
\end{figure}

Figure\,\ref{Fig:final_cbd} illustrates the computed CBD values for the entire sample. On the y-axis, we depict the IB224 magnitude. We selected this specific band to construct the diagram due to its central position within the SED and its independence from the influence of CO band depth. Furthermore, the data in this band exhibit high quality, with fewer emission lines compared to the IB206-band, which corresponds to the data with the highest quality. In the left panel, typical errors for CBD values in bins of 0.5 magnitudes are displayed. The errors are adjusted as we explain in detail in Appendix \ref{CBD_errors}.

\begin{figure*}[!ht]
\centering
\includegraphics[width=\textwidth]{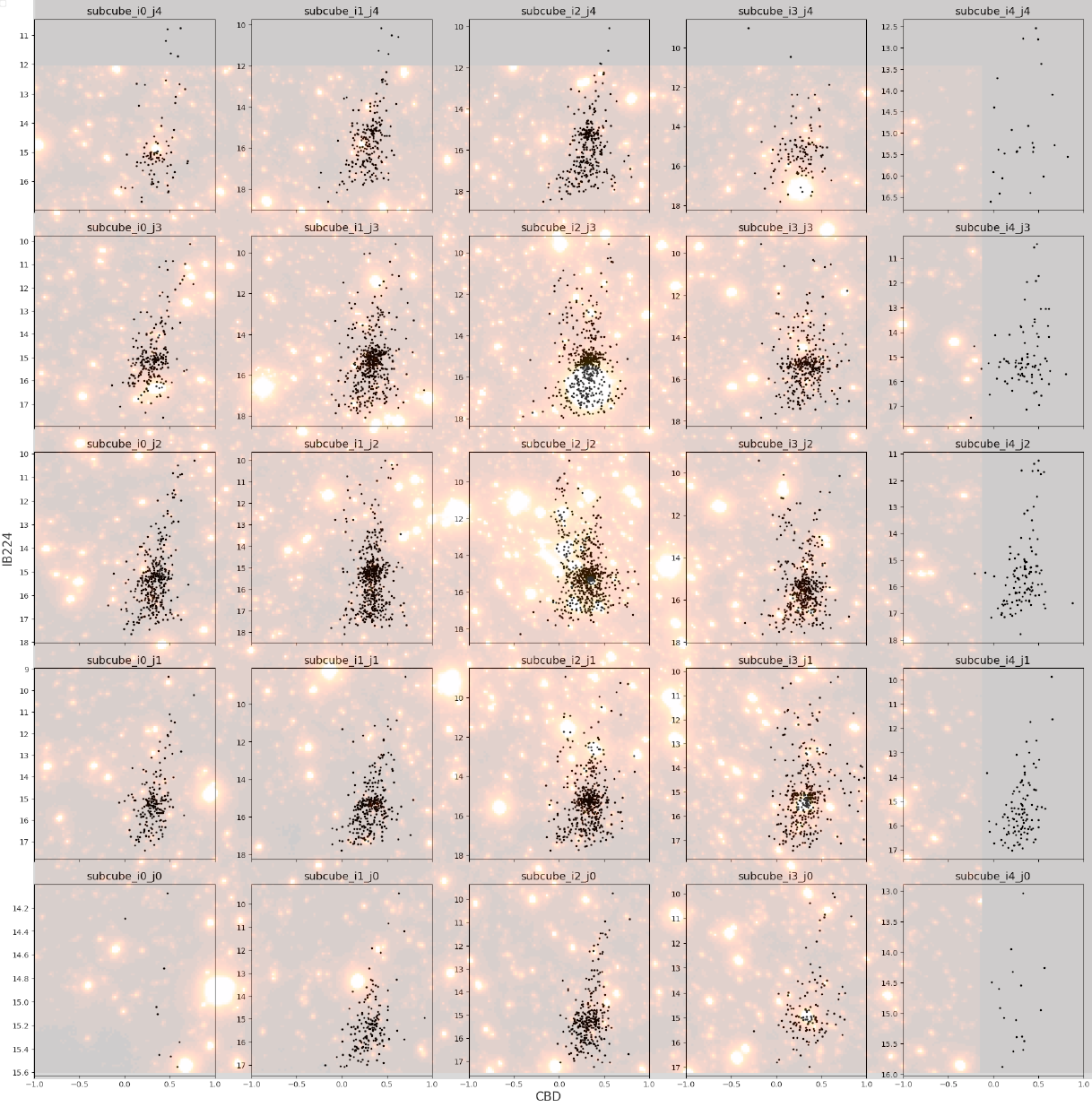}
\caption{CBD diagrams along the FoV. The size of each sub-field is around $8.6''$ x $8.6''$ size. The panels corresponding to the different sub-fields overlay the specific sub-field in the background image, corresponding to the epoch on June 12, 2004.}
  \label{Fig:cbd_fov}
\end{figure*}

We studied the variation of the CBD diagrams across the FoV. Figure\,\ref{Fig:cbd_fov} displays the CBD values within the different subcubes into which we previously divided the FoV, as described in Section\,\ref{sec:local_calibration}. We can observe that in the regions closer to the centre, a branch of bright young stars (small value of CBD) is more prominent compared to the outer regions as we expected.

\subsection{Cross matching with spectroscopic studies} \label{sec:match}
We conducted a comprehensive comparison with several works from the literature. The following studies were included in our analysis: \cite{do2013densities,do15metallicy,feldmeier2014isaac,feldmeier2015kmos,feldmeier2017kmos,feldmeier2020asymetry,fellenberg2022young,fritz2016mass,gillessen2009orbits,gillessen2017orbits,habibi2019late,stostad2015types,yelda2014clockwise}. We excluded \cite{feldmeier2014isaac} from our analysis due to its low spatial resolution, which posed challenges when attempting to match it with our high-resolution data. Moreover, all the stars mentioned in \cite{feldmeier2014isaac} were accounted for either in \citet{feldmeier2015kmos,feldmeier2017kmos}.

We exclusively considered stars with positions not based on orbital information. Hence, our input catalogs do not include the stars S175 (early), and S145 (late) \citep{gillessen2017orbits,fellenberg2022young}. The primary source for position data in the central MPE dataset is generally \cite{gillessen2009orbits} because it provides more polynomial fits of position against time and offers the advantage of an early zero point. This is particularly useful since our data was acquired in 2004. We obtained the position of S2 in 2004 using Fig.~13 from \cite{gillessen2009orbits}. For the remaining stars, we employed a polynomial fit to determine their position in 2004.5. However, we updated the star types using the type tables from \cite{gillessen2017orbits} and \cite{habibi2019late} to ensure the most current information. All stars from \cite{fritz2016mass} with radial velocities are considered late type.

We considered matches as stars coinciding within $0.1''$ in position and $0.76$ magnitudes in brightness (a factor of 2 in flux). In cases of ambiguity, we selected the nearest neighbour, correcting for any systematic offsets between the used catalogues and our data using median offsets and an iterative procedure. 

We compared the classifications obtained using this method. We observed occasional discrepancies in classifications from \cite{fritz2016mass}, likely due to their fully automated procedure. Consequently, we excluded stars classified solely as late-type in \cite{fritz2016mass}, except those used in \citet{pfuhl2011SFH}, as the latter underwent rigorous analysis, ensuring their late-type classification certainty. The resulting matched list still includes five stars with conflicting spectral types from different sources. We classified these stars using the CBD diagram, where they are clearly distinguishable and exhibit magnitudes brighter than 14. In all but one conflicting case, they were classified as early-type stars.

As a final step, we compared the spectroscopic classifications with ours used in Section\,\ref{sec:photo_calibration}, which led to the inclusion of a few more stars but also revealed some discrepancies. We re-examined the early and late stars from \cite{feldmeier2017kmos}, which have been assigned to different catalogue stars using different methods. The correct match was always identifiable for early-type stars. However, for late-type stars, in some cases, the \cite{feldmeier2017kmos} classification likely resulted from two similar bright stars merging in the seeing-limited spectrum. In these cases, neither star was definitively assigned a certain type, and they are excluded from both the early and late samples in the following sections. Further details on interesting stars can be found in Appendix\,\ref{star_details}. Figure\,\ref{Fig:final_cbd} shows the CBD diagram for all spectroscopically classified stars: 982 late-type stars (red points) and 212 early-type stars (blue points). We explored also some interesting stars in Appendix\,\ref{star_details}. 

\subsection{Identification of foreground and background stars} \label{sec:fore}

Before applying the classification methods, we eliminated foreground and background stars from our final list. Our method for detecting foreground and background stars is based on the methodology outlined by \cite{buchholz2009composition}, with several refinements. Specifically, we employed the H-IB224 colour for extinction determination. To calibrate, we adopted the extinction values determined by \citet{buchholz2009composition}, adjusted to the scale of the new extinction law \citep{schodel2010peering, fritz2011line} using a scaling factor of 0.84. In addition to the H-IB224 colour, we incorporated the CBD value as the second parameter to account for the changing temperature in a linear way, complementing the information provided by the H-IB224 colour in our analysis.
For stars lacking an H-band measurement, particularly prevalent at the lower end outside the H-band image coverage, we used the linear fit slope $m$ and recalibrated using the same methodology.
The lower H-IB224 colour cut for excluding foreground stars is set at $1.68$, aligning with the value used by \citet{buchholz2009composition} and translated into our extinction system.
Opting for a more inclusive cut results in a noticeable concentration of foreground stars, a phenomenon not anticipated, and this adjustment does not contribute additional early-type stars to the foreground classification.
By analogy, we designated stars with a colour greater than 4.2 as background. Additionally, we included star 5251 (IRS34) in the background category, as it distinctly aligns with the red population of bright stars. 
This results in a total of 37 foreground and 22 background stars, predominantly excluded from subsequent analyses.

We excluded foreground and background stars primarily due to the magnitude-dependent nature of the CBD value. The varying distances to these stars lead to the conclusion that a different model is more suitable for them than for GC stars, which are our primary targets. Although most of the bright stars (IB224 < 12) with high extinction are likely GC members, we still excluded them, as very few unknown stars share the combination of being both bright and red. Furthermore, many of these bright and red stars are of known types\footnote{Except our Id 1573, which is identified as a late Mira \citep{feldmeier2017kmos}, and IRS2L, for which a specific type could not be determined despite the availability of numerous spectra, as discussed in \cite{feldmeier2015kmos}, the majority of the stars are early types.}.
Consequently, our final sample comprises 201 early and 980 late stars with spectral types, which are likely to be GC stars and exhibit minimal reddening. 

\subsection{Preliminary classification} \label{sec:preliminar_class}
In this section, we conduct a preliminary classification of the stars following the method of \cite{nishiyama2023search}. They used also narrow-band filters to identify potential hot, massive stars, and intermediate-age stars in the GC. 

\begin{figure} [ht!]
\centering
\includegraphics[width=\columnwidth]{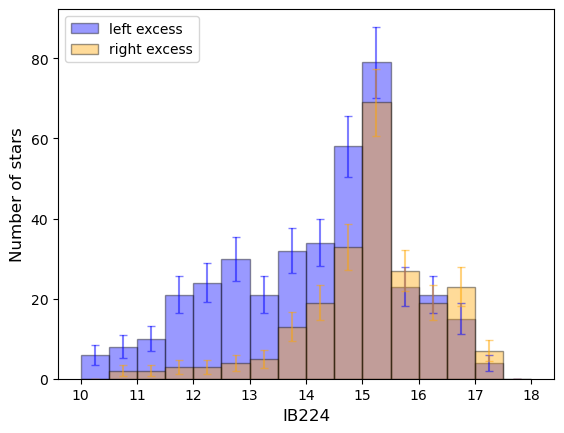}
\caption{CBD histograms for the excess of stars, with CBD < 2$\sigma$ (left excess), and with CBD > 2$\sigma$ (right excess).}
  \label{Fig:excess}
\end{figure}

\begin{figure} [ht!]
\centering
\includegraphics[width=\columnwidth]{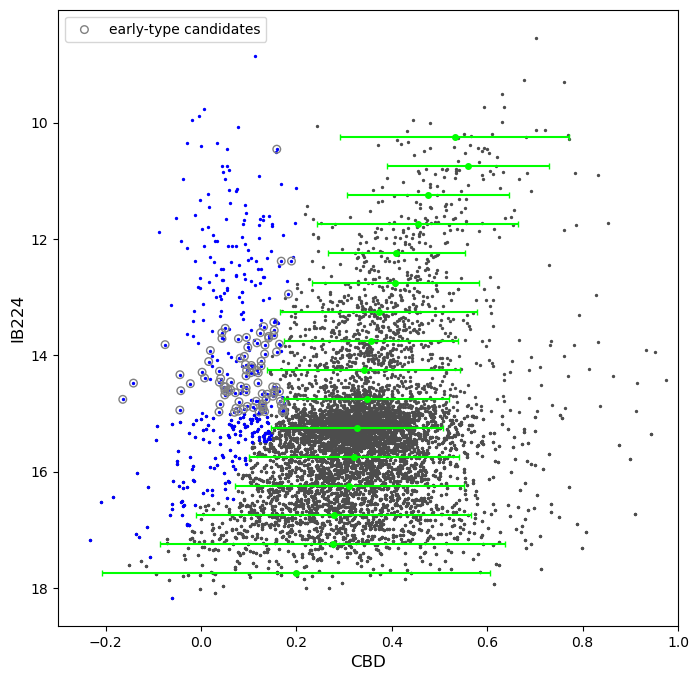}
\caption{CBD diagram of the sample. The green horizontal bars represent $\pm 2\sigma$, where $\sigma$ stands for the standard deviation of CBD within each magnitude bin. Stars with CBD values below $2\sigma$ are categorized as young candidates and denoted by blue data points in the diagram. Limiting the sample to stars brighter than 15 magnitudes, and focusing solely on the new candidates not previously spectroscopically identified, early-type candidates are represented by grey circles.}

  \label{Fig:nishiyama_CBD}
\end{figure}

We selected stars with magnitude errors in IB224 < 0.1, and with adjusted CBD error < 0.2. We created CBD histograms for stars within 0.5 magnitude wide bins in IB224. The majority of stars at all magnitudes lie on the Red Giant Branch (RGB). We fitted each histogram with Gaussian functions to extract the mean magnitude $\mu$ and the standard deviation $\sigma$ of the RGB CBD values in each bin. For IB224$<15$ mag, we considered the stars with a CBD value of $0.15$ to be centred on the RGB. Subsequently, we computed the excesses of stars with CBD < 2$\sigma$ (left excess) and with CBD > 2$\sigma$ (right excess) in each magnitude bin. Figure\,\ref{Fig:excess} shows that there is a clear excess of stars with low CBD values for magnitudes brighter than $15$. In our final classification, we considered bins where the left excess count was more than double the right excess count. This ensured a probability of over 50\% for stars in these bins to be classified as early-type.
We identified 248 stars with a likelihood exceeding 50\% of being early-type, 93 of which are new candidates. Additionally, 133 have been spectroscopically confirmed as early-type stars, while 22 are misclassified due to their spectroscopic identification as late-type stars. Figure\,\ref{Fig:nishiyama_CBD} shows in blue points all stars corresponding to the left excess with magnitudes < 15. New early candidates are represented by grey circles in the diagram. 

When comparing the result of this simple classification with the sample of spectroscopically classified stars with IB224 < 15, we find that we have successfully identified 82.1\% of the known early-type stars and 96.5\% of the late-type stars. Although this preliminary classification is a good first step, it does not provide the probabilities which are useful for exploratory analysis. In the next sections, we explore three classification methods to provide probabilities of being early-type for the list of stars.

\section{Bayesian inference}\label{sec:bayesian}

After the initial approximate classification, we employed a more refined Bayesian analysis. We fitted models to the two branches in the CBD diagram (CBDD) corresponding to late and early-type stars (see Fig. \ref{Fig:s_2comp_CBD}). Subsequently, we inferred the probabilities of each star belonging to either the early-type or late-type branch. Throughout the following sections, we refer to the parts of the CBDD that predominantly contain early or late-type stars as the `early-type branch' and `late-type branch', respectively. The prior is determined by the initially estimated density of spectroscopically classified GC stars in the respective branches. 

In the initial step, we conducted a separate analysis of the properties of known late and early-type stars. This separation was essential due to significant differences in the available literature data for early and late stars in the GC. The data were acquired using diverse instruments, observational methods, setups, and analysis techniques. We fitted separate models to the two branches. The results obtained from these two models enabled us to assess the probability that a star is of the early-type category based on observed parameters such as IB224 magnitude, CBD and its uncertainty. 

The second step in our methodology involved refining the prior probability distribution. In the first step, we considered a flat or uniform prior probability distribution with a value of 0.5. However, recognizing the limitations of a flat prior, we adopted an iterative approach to enhance its accuracy. This refinement was achieved through a function that considered both IB224 magnitude and the distance from Sgr~A*.

\subsection{Likelihood fits}\label{sec:lik_fitting}

We used the \texttt{PyMC3}\footnote{\url{https://www.pymc.io/}} package to fit models to the early and late-type branches in the CBDD separately. We assumed a normal distribution of the uncertainties as our baseline. We treated the standard deviation of the data points relative to the model as a free parameter, which was added in quadrature to the CBD errors.  This was necessary because the CBD uncertainties alone were insufficient to account for the spread observed in the CBD values.

We found that a second-order polynomial function of magnitude led to the best results to fit the early\footnote{For the early-type branches, the significance of the quadratic term is not always substantial; in some cases, the best fit is close to linear. We retained the quadratic term to maintain consistency in the model structure. In practice, the impact of the quadratic term can be minimal.} and late-type branches. We subtracted the value of 14 from all magnitudes to reduce the correlations between the different polynomial parameters.

Our base model is the normal distribution, but we frequently expanded it by incorporating additional parameters to accommodate deviations. We employed two distinct variants:

\color{black}
\begin{itemize}
\item Student-t Distribution: This symmetric distribution has fatter tails than the normal distribution, which can be useful when a data set has a stronger kurtosis than the latter. 
\item An additional component with a constant offset. This option is primarily valuable for distributions with significant skewness and requires two additional parameters.
\end{itemize}

We employed wide priors when fitting the model parameters, ensuring that their impact on the relevant results remained negligible\footnote{The exception is when the curve corresponding to the model for the early branch crosses the curve from the model corresponding to the late branch. In such cases, we used the CBD value and uncertainty at the faint end of late as a prior for the CBD value of the late-type fit in that region.}.

After the initial fit with the normal model, we utilized the best-fit parameters to normalize the CBD values. Subsequently, we calculated the kurtosis and skewness of the distribution. Both had values often exceeding $2.5$ and, in many cases, even 5 or more, especially in the case of the kurtosis for the late branch. In terms of sign, the kurtosis was consistently positive, as expected when outliers play a significant role. The skewness typically showed a positive direction for early-type stars and a negative one for late-type stars. 

Subsequently, we predominantly employed a Student-t distribution for both early-type and late-type stars. The choice for late-type stars was evident due to the highly significant kurtosis, while for early-type stars, the significance of skewness was approximately 1 sigma larger. However, attempts to address this by using an asymmetric distribution for early-type stars resulted in implausible early-star candidates within the CBD diagram. Consequently, we opted for a Student-t distribution for early-type stars as well. This choice is justified by the absence of a clear mechanism causing asymmetry in the early CBD diagram, while a Student-t distribution can account for underestimated errors in certain stars. While we explored alternative distributions for early-type stars, our findings indicate that the specific choice of model for early-type stars did not significantly impact the results. In contrast, the model selection for late-type stars is more crucial, as the influence of the late model becomes more pronounced in the wings of the distribution, given the larger number of late-type stars.

\begin{figure} [ht!]
\centering
\includegraphics[width=\columnwidth]{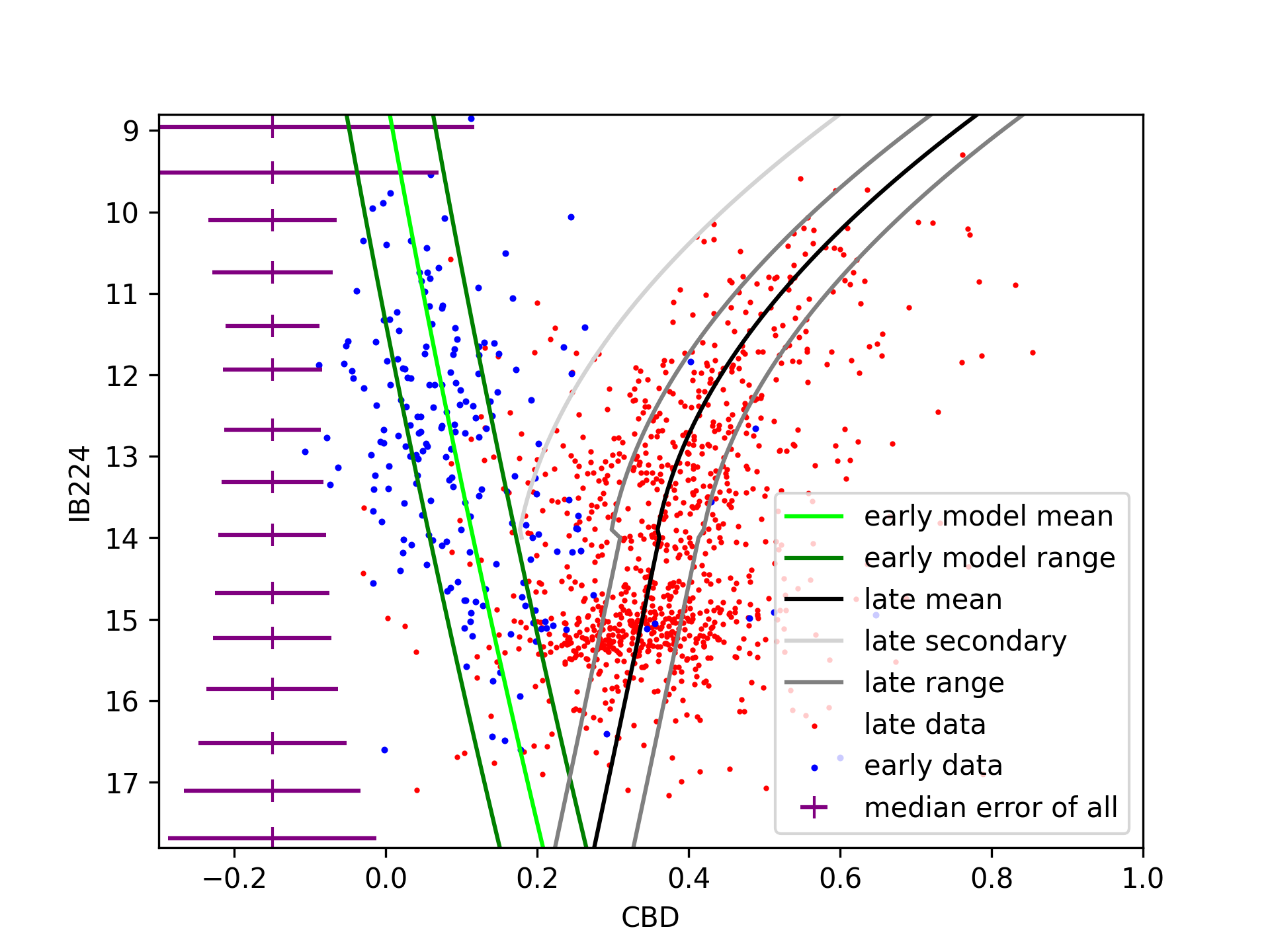}
\caption{Bayesian student-t fit for early-type stars and a two-component model fit for late-type stars. The latter includes sub-models for stars brighter and fainter than 14 magnitudes, using a two-component normal model with a quadratic mean equation and a student-t model with a linear mean equation, respectively. The left side of the figure displays median errors for all stars in constant bins (0.65 mag), with lines representing the mean and 1-$\sigma$ range of the models.
}
  \label{Fig:s_2comp_CBD}
\end{figure}

We further investigated the possibility of employing two Student-t distributions for late-type stars, considering the impact of late-type stars with a warmer than average effective temperature \citep{pfuhl2011SFH,do15metallicy,feldmeier2017kmos}. We identified a second component shifted by $0.2$ magnitudes towards redder CBD values, suggesting the presence of low metallicity stars (see section\,\ref{sec:discussion}). This component corresponds to $4.4^{+1.9}_{-1.4}$\% of all late-type stars, consistent with the literature. \cite{pfuhl2011SFH} reported a fraction of 10\%, while \cite{do15metallicy} and \cite{feldmeier2017kmos} reported fractions of 6\% and 5.2\%, respectively.
From the data, it is apparent that we underestimate the second component. This suggests that the second component cannot be attributed to the RC due to its distinct characteristics. In response, we explored models that divide the late stars by magnitude at 14, introducing a second component only for the brighter portion. Consequently, our primary model incorporated a two-component normal distribution with a quadratic mean model for stars brighter than 14 magnitudes and a student-t distribution with a linear mean model for fainter stars.
This results in a warm star fraction of $8.9^{+5.6}_{-3.5}$\% consistent with the literature. Figure~\ref{Fig:s_2comp_CBD} show the best fits.

\subsection{Iteration on priors}\label{sec:it_prior}

\begin{figure} [ht!]
\centering
\includegraphics[width=\columnwidth]{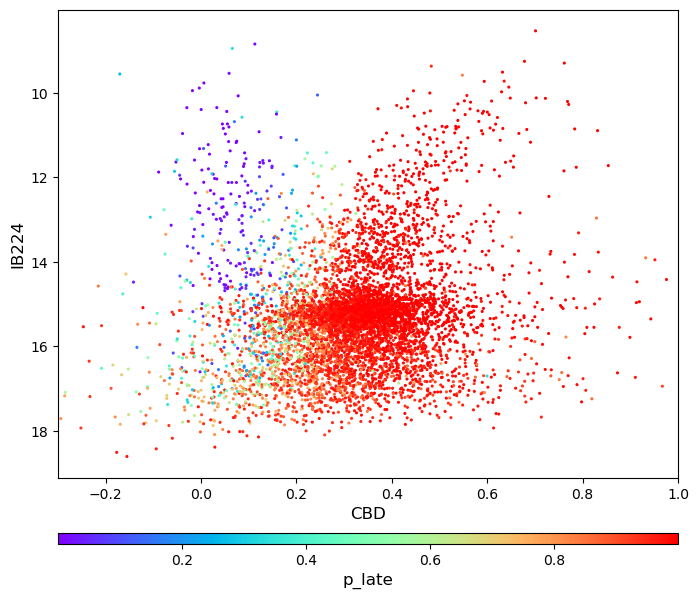} 
\caption{Final CBD diagram with the posterior distribution for our favoured model, employing a student-t distribution for early-type stars and for late-type stars with a magnitude split (two component normal distributions for objects brighter than 14 magnitudes, alongside a Student-t distribution for objects fainter than 14 magnitudes) and integrating a prior.
}
  \label{Fig:bayesian_cbd_post}
\end{figure}

In the second step, we refined the prior. Without additional information, the prior would assume equal probability for early and late, which leads to an overestimation of early-type stars in contrast to the real situation at the GC, where there are more late-type stars than early-type stars. To enhance the prior with information from spectroscopically classified stars, we incorporated the KLF using the IB224 magnitude and the radial distance from Sgr~A* in the calculation. We binned the data based on these properties and utilized the \texttt{curve\_fit} tool for fitting. To ensure completeness, we limited our analysis to stars brighter than 16 magnitudes and imposed a radial limit of 14 arcseconds to mitigate incompleteness effects.
We also excluded the areas of the image which were observed by two or fewer exposures. To do this, we initiated the process with probabilities derived from the flat prior case. In subsequent steps, we substituted the probabilities with specific early or late classifications when available. This replacement was necessary to ensure convergence.

We modelled the KLF as follows:

\begin{equation}
\label{eqn:prior}
f(\rm{x})=[a\cdot e^{b\,x}+h\cdot e^{-(\rm{x}-\mu)^2/\rm{\sigma}^2}]\cdot[1-1/(1+e^{-c\cdot (x-d)})],
\end{equation}

where $x$ is the magnitude, $h$ is the amplitude which determines the peak value of the Gaussian, $\sigma$ is the standard deviation, and $\mu$ is the mean value.
We employed an exponential function as the baseline, complemented by a Gaussian component to characterize the RC. This methodology aligns with the approach employed in \cite{schodel2010accurate}. The multiplication of the $c$ and $d$ parameters modelled the deviation from the exponential function, which for our magnitude limit 16.0 occurs at the bright end. For early-type, the Gaussian is always omitted since there is no blue clump.
For the radial profile, we used the Nuker profile, as described in \cite{fritz2016mass}. For early stars, the preferred value for $\alpha$, determining the sharpness of the transition, was often very high (indicating essentially a broken power law), which led to numeric fitting problems. In response, we fixed $\alpha$ at 64, which is nearly broken but does not present numeric problems.

The prior is defined by the following equation, defining the probability for early-type as:

\begin{equation}
\label{eqn:prob_e2}
p_{\textit{e}} =
\frac{d(\textit{e}, \textit{IB}) \cdot d(\textit{e}, \textit{r})}{<d(\textit{e}, \textit{IB})>}.
\end{equation}

It utilizes the density at a given magnitude and radius of the stars, which is then divided by the density at the mean radius of the early-type population. Including this last term is necessary because the magnitude density is obtained at this radius, making the radial term relative to it. 

We applied the same procedure to define the prior probability for late-type stars, as follows:

\begin{equation}
\label{eqn:prob_l2}
p_{\textit{l}} =
\frac{d(\textit{l}, \textit{IB}) \cdot d(\textit{l}, \textit{r})}{<d(\textit{l}, \textit{IB})>}.
\end{equation}

To normalize the total probability to one, we employed the following scaling: 
\begin{equation}
\label{eqn:prob_e_norm}
p_{\textit{e-norm}} =
\frac{p_{\textit{e}}}{p_{\textit{e}}+p_{\textit{l}}}.
\end{equation}

This process was repeated multiple times until all changes become very small, typically around 36 iterations.

We present our final priors for several example radii and magnitudes in Table~\ref{Tab:priors}. Notably, a pronounced radial trend is evident, accompanied by a comparatively weaker trend in magnitude. The prior bias towards late stars in terms of magnitude aligns with expectations, particularly evident at the RC.

\begin{table}
\centering
\caption{Prior distributions employed for Bayesian inference across various magnitudes and distances. }
\label{Tab:priors} 
\begin{tabular}{llll}
\hline
\hline
magnitude & $R = 1''$  & $R = 10''$  & $R = 20''$   \\
\hline
10 &   0.138 &  0.985 &  8.015\\
12 &  0.222 &  1.577 & 12.827\\
14 &  0.429 &  3.056 & 24.864\\
15.25 & 2.59 &  18.35 & 149.971\\
16.5 &  0.829 &  5.900 &  48.0\\

\hline
\end{tabular}
\tablefoot{
The values of the priors represent the ratio of late stars to early stars.
}
 \end{table}

Figure\,\ref{Fig:bayesian_cbd_post} displays the posterior (the product of the prior and the likelihood) of our preferred model. Figure~\ref{Fig:mag_r_split} shows the posterior of the distribution of stars as a function of magnitudes. Overall, 6.0\% of all spectroscopically classified stars without extinction flags are misidentified in terms of magnitudes and radii. 
The logarithmic loss among them is 0.1610. Integrated overall probabilities excluding stars with spectral types, we find a total of 451.3 early-type stars, as shown in Table~\ref{Tab:sum}. We define an early-type candidate as a star with p(early)$>0.5$ and without a spectroscopic classification in our input catalogue. We identified 204 early-type candidates (see Table~\ref{Tab:can}). 

\section{Machine learning methods }\label{sec:ML}
In this section, we explore two machine-learning approaches for star classification: a Multi-layer perceptron (MLP) and Gradient Boosted Trees (GBT). It is important to note that while Bayesian analysis provides interpretability and a probabilistic framework for handling uncertainty, its performance for fainter stars is heavily dependent on having a prior derived from unbiased data. Choosing incorrect priors may introduce bias into the results, therefore, the effectiveness of Bayesian analysis for fainter stars depends on having a well-informed, unbiased prior that aligns with the true distribution of star types, ensuring a more accurate and reliable classification. In contrast, the primary strength of MLP and GBT lies in their capability to discern intricate patterns within data. Although they can be prone to overfitting, we have implemented mechanisms in both methods to prevent this issue, as we explain in the following sections. Moreover, all the methods depend on the spectroscopic sample.

Both models are built upon a dataset consisting of 1181 stars with spectroscopic classifications. Among these, there are 201 early-type stars and 980 late-type stars (see Section\,\ref{sec:match} and Section\,\ref{sec:fore} ).

\subsection{Multi-layer perceptron}\label{sec:MLP}
An MLP is a type of artificial neural network designed for supervised learning. We used the Python-based \texttt{Tensorflow}\footnote{\url{https://www.tensorflow.org/}}(TF) framework. The neural network model is built using the Keras Sequential API (Application Programming Interface). We defined a simple MLP architecture with an input layer, a dropout layer, a hidden layer, and an output layer. The model is suitable for tasks such as classification. The classification is based on five features: the magnitudes of the stars in the IB224 filter, CBD values, their associated errors, and the projected distance to Sgr~A*. 

Initially, we partitioned the dataset into training (90\%) and testing (10\%) sets. An important preprocessing step involved standardization to normalize the features. This is particularly beneficial for machine learning algorithms, especially those dependent on distance measures or gradient-based optimization. Given that our features have diverse characteristics, such as magnitudes and distances, we adopted different scalers. We used \texttt{scikit-learn}\footnote{\url{https://scikit-learn.org/}}tool \citep{scikit-learn}. Specifically, we used \texttt{StandardScaler} for the magnitude and CBD subsets, ensuring a mean of 0 and a standard deviation of 1, and \texttt{MinMaxScaler} for the subset representing distances. The latter scales data to a specified range, typically [0, 1], adjusted to the radial distances to avoid negative values.

To augment our training dataset and minimize the risk of the model memorizing specific patterns, we generated additional training samples by doubling their final number. This process included introducing random noise into the features and shuffling the data to randomly alter the order of rows. Shuffling is a crucial step during training to prevent the model from learning patterns based on the sequential arrangement of the data.

Our MLP model was constructed with the following architecture: an input layer comprising 5 neurons with `Exponential Linear Unit' (ELU) activation, followed by a dropout layer with a 50\% dropout rate. The subsequent hidden layer consists of 5 neurons using `Rectified Linear Unit' (ReLU) activation. The final layer, designed for binary classification distinguishing late- and early-type stars, comprises 1 neuron with a `Sigmoid' activation function. The model incorporates 66 trainable parameters. For optimization, the `Adam' (Adaptive Moment Estimation) optimizer is employed, utilizing a binary `binary\_crossentropy' loss function designed for binary classification tasks. In our training configuration, we set the learning rate to 0.001, a critical parameter governing the magnitude of weight adjustments during training. To mitigate overfitting, we incorporated an early stopping callback with a patience of 4. During the model training process, we allocated 20\% of the training data for validation purposes. The neural network completed 44 training epochs. The validation accuracy achieved 93.9\%, with a false positive rate of 6.1\% on the validation set. When evaluated on the testing dataset, the model demonstrated an accuracy of 95.8\%, with a false positive rate of 4.2\%.

We also considered the use of class weights during training to address the imbalance in our dataset, which consisted of approximately 83\% more late-type stars than early-type stars. However, adopting this approach resulted in an excessive number of false negatives for the young-type candidates, making the results unreadable and challenging to interpret. The imbalance stems from the inherent prevalence of late-type stars in the region we analyzed. To maintain interpretability, we opted not to use class weights in our final model. We acknowledge the impact of this imbalance on our training data and discuss its implications for result interpretation.

\begin{figure} [ht!]
\includegraphics[width=\columnwidth]{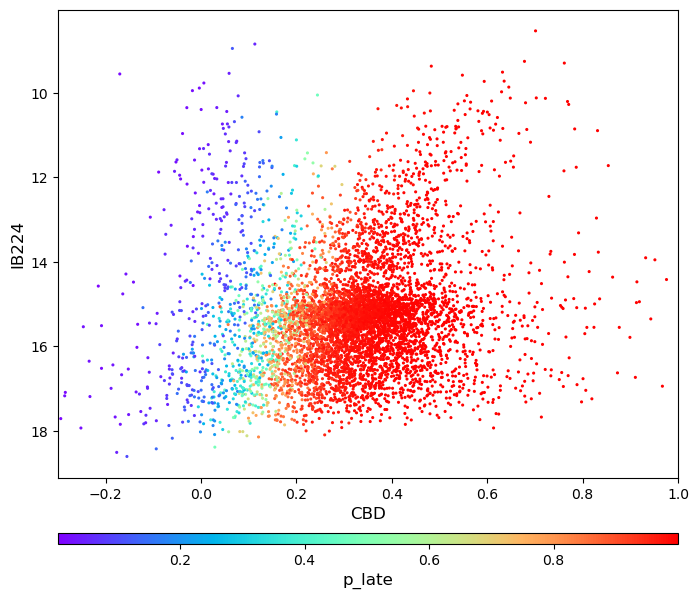}
\caption{CBD diagram for the entire sample, where the colours of the points indicate the probabilities predicted with the MLP method.}
  \label{Fig:CBD_nn}
\end{figure}

We applied this model to the IB data, predicting the probabilities of stars being early-type. Figure\,\ref{Fig:CBD_nn} shows the final CBDD for the entire sample. We find 498 likely new early-type stars (p(early)$>0.5$) (see Table~\ref{Tab:can}). Integrated over all probabilities, we find 641.5 early-type stars (see Table~\ref{Tab:sum}) excluding stars with types.
In Appendix \ref{MLP_results}, we provide additional details about the procedure. Overall, 6.3\% of all spectroscopically classified stars are misidentified. The logarithmic loss of them is 0.1710. 

\subsection{Gradient boosted trees}\label{sec:GBT}

We employed the \texttt{XGBoost}\footnote{\url{https://xgboost.readthedocs.io/}} \citep{chen2016xgboost} algorithm, which is a fast gradient boosted trees algorithms. We experimented with various feature choices, including incorporating multiple magnitudes as done in \cite{plewa2018random}. However, we determined that the most effective approach is to utilize only the CBD and a single magnitude. Our ultimate feature selection comprises IB224, CBD, CBD errors, IB224 errors, and the radius. Due to the restricted size of the training sample, we implement stratified cross-fold validation, leaving out one-fifth of the data in each iteration.

We primarily utilized the standard settings of \texttt{XGBoost}; however, to prevent overfitting, we explored all possible max-depth values and steps of $\sqrt{2}$ in the alpha-regularization, covering a broad range to ensure finding the optimum. Our optimal solution was achieved with a max-depth of 2 and a regularization of 2.04. 

In the application, we used the average of all five folds\footnote{In five-fold cross-validation, the dataset is split into five parts. The model is trained and evaluated five times, each time using a different part for evaluation and the rest for training. This helps ensure a robust assessment of the model's performance across different subsets of the data.} for stars which were not used in the fitting process. For the other stars, we use the average of the folds in which the star was in the training sets, which avoids overfitting since we use the test fold for the parameter optimization. We show the CBDD with the final probabilities obtained in Fig.~\ref{Fig:rf_cbd}. In the Appendix\,\ref{GBT_results}, Figure~\ref{Fig:rf_plots} shows the probabilities of the stars as a function of the radius. We find 311 likely new early-type star candidates (see Table~\ref{Tab:can}). Integrated over all probabilities, we find 463.9 new early-type stars (see Table~\ref{Tab:sum}. Overall, 5.8\% of all spectroscopically classified stars are misidentified. The logarithmic loss of them is 0.167.

\begin{figure} [ht!]
\centering
\includegraphics[width=\columnwidth]{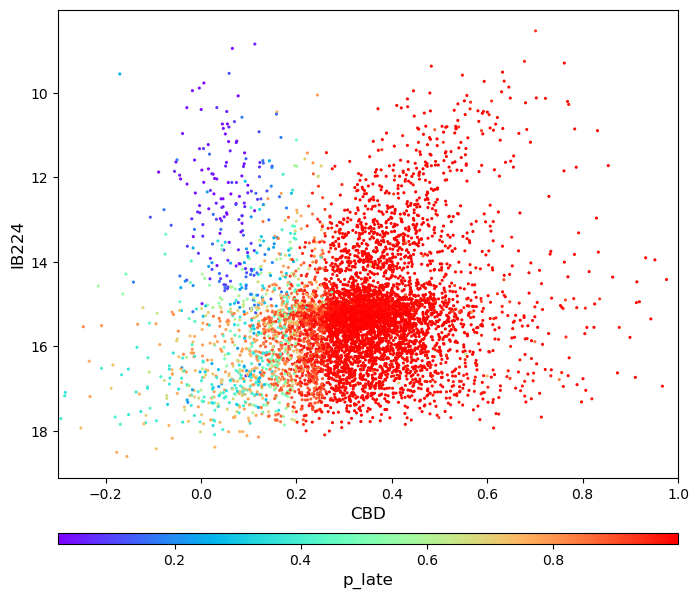} 
\caption{CBD diagram for the entire sample, where the colours of the points indicate the probabilities predicted with the XGBoost method. 
}
  \label{Fig:rf_cbd}
\end{figure}

\section{Results}\label{sec:result}

\subsection{Early-type and late-type star candidates}\label{sec:results_classification}
We summarize the early-type candidates identified by our classification methods as well as for a radial distance ($R_{cut}$=14$''$) and IB224 magnitude ($m_{cut}$=16 mag) in Table~\ref{Tab:can}.

\begin{table}
\centering
\caption{Number of new early-type candidates.}
\label{Tab:can} 
\begin{tabular}{lllll}
\hline
\hline
Method & all & $m<16$ & $R<14''$ & $m<16$ \& $R<14''$ \\
\hline
MLP & 498 & 229 & 351 & 148  \\
XGBoost & 311 & 141 & 294 & 124  \\
Bayesian & 204 & 134 & 185 & 116  \\
Common & 155 & 98 & 145 & 88  \\
\hline
\end{tabular}
\tablefoot{
We defined a candidate early-type star as one with an unknown spectral type and a probability of being early larger than 0.5. The Common row includes the early-type candidates identified consistently by all three methods. The magnitudes $m$ are extinction-corrected IB224 magnitudes after the local calibration (see Section\,\ref{sec:local_calibration}). $R$ is the projected distance from Sgr\,A*.
}

 \end{table}
There are noticeable differences, with the counts decreasing from MLP to XGBoost and further to Bayesian. Approximately 3/4 of the stars classified as early by the Bayesian method are also classified as early by the other two, indicating a significant overlap. 
Limiting the sample to brighter stars and those closer to Sgr\,A* helps to minimize differences between the methods. When considering only stars within $R=14''$ and brighter than 16 magnitudes, the numbers of candidates differ by less than 22\%. This is to be expected because the spectroscopically classified stars are bright and concentrated around Sgr\,A*. Also, the depth of the CO bandhead absorption decreases with magnitude because the fainter stars will be hotter giants \citep[see, e.g., Fig.\,1 in][]{do2013densities}. 

\begin{figure*}[!ht]
\centering
\includegraphics[width=\textwidth]{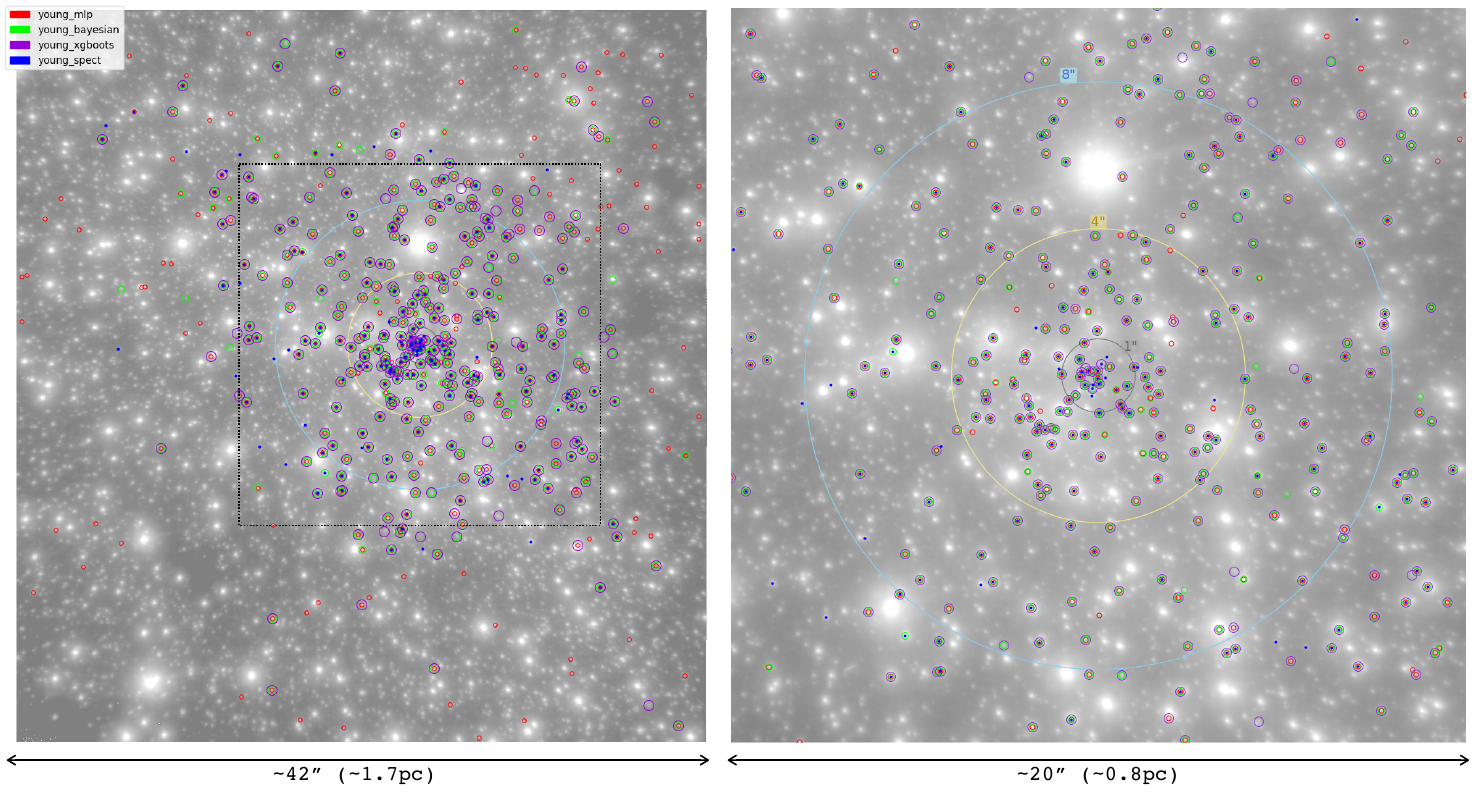}
\caption{Map illustrating the positions of spectroscopically classified stars (blue points) and early-type candidates brighter than 16 magnitudes for the different methods, superimposed on the image from 12 June 2004. The three distinct regions, as identified in \cite{fellenberg2022young}, are delineated by circles in the images. The right hand panel shows a zoom into the central 20$''$, marked by the black square on the left.}
  \label{Fig:map_candidates}
\end{figure*}

Figure \ref{Fig:map_candidates} illustrates the positions of young candidates brighter than 16 magnitudes, detected using the MLP (red circles), Bayesian (green circles), and XGBoost method (violet circles). We discuss some especially interesting candidates in Appendix~\ref{star_details}.

Finally, we have also examined the late-type candidates with p(late)>0.5 that lacked prior spectroscopic classification. Out of all the methods utilized, a total of 4806 common late-type candidates were identified. Specifically, employing the MLP method led to the discovery of 4852 new late-type candidates, while the XGBoost method revealed 5039, and the Bayesian approach uncovered 5146. It is important to emphasize that while late-type star classification may not be as groundbreaking as early-type classification, its significance lies in the fact that late-type candidates typically exhibit prominent CO absorption bands, making their classification relatively straightforward even in the outer regions. Additionally, the higher number of late-type stars makes the identification also more robust causing that of the spectroscopically classified stars clearly more are misidentified among the early (~17\%) than among the late (~4\%) stars. Consequently, stars classified as late-type by our method have a high likelihood of being accurate, and thus, unlikely to be early-type.

\subsection{Structure of the cluster}\label{sec:results_structure}

Instead of using candidates identified by a probability cut, the probabilities can be used directly to infer quantities of interest, such as luminosity functions and radial density distribution of the different types of stars. We present the sum of the early-type probabilities in Table~\ref{Tab:sum}, excluding stars with known spectral types so that the differences between the different classification methods are more evident.

\begin{table}
\centering
\caption{Summed Probability of new early-type candidates.}
\label{Tab:sum} 
\begin{tabular}{lllll}
\hline
\hline
Method & all & $m<16$ & $R<14''$ & $m<16$ \& $R<14''$ \\
\hline
MLP & 641.5 &  338.9 & 438.6  & 219.8  \\
XGBoost & 463.9 & 237.9 & 353.0 & 169.8  \\
Bayesian &  451.3 & 233.3 & 380.5 & 188.2  \\
\hline
\end{tabular}
\tablefoot{
Here the probabilities of stars are summed up, excluding stars with known spectral types. 
The magnitudes $m$ are extinction-corrected IB224 magnitudes after the local calibration (see Section\,\ref{sec:local_calibration}). 
}
 \end{table}

As can be seen, when considering probabilities the differences between the classification methods are smaller than when merely using counts of classified stars. In Table~\ref{Tab:can}, we observe that the MLP method produced 2.4 times as many candidates as the Bayesian method. For probabilities, the factor is reduced (1.4). At larger $R$ the two machine-learning approaches, particularly the MLP method, show a significant excess of the summed early-type probabilities over the Bayesian methodology. This may be related to the Bayesian prior that imposes a decrease of early-type stars with $R$. 

\begin{figure} [ht!]
\centering
\includegraphics[width=\columnwidth]{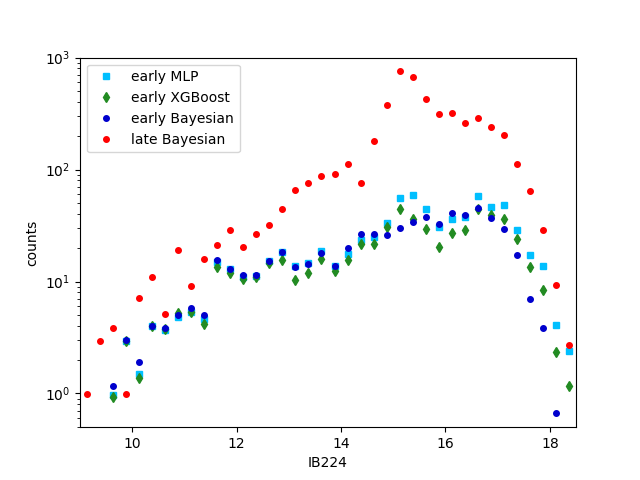} 
\caption{Luminosity functions obtained with the different methods summing up probabilities. For the dominating late component, the different algorithms obtain very similar results.
}
  \label{Fig:klf1}
\end{figure}

\begin{figure} [ht!]
\centering
\includegraphics[width=\columnwidth]{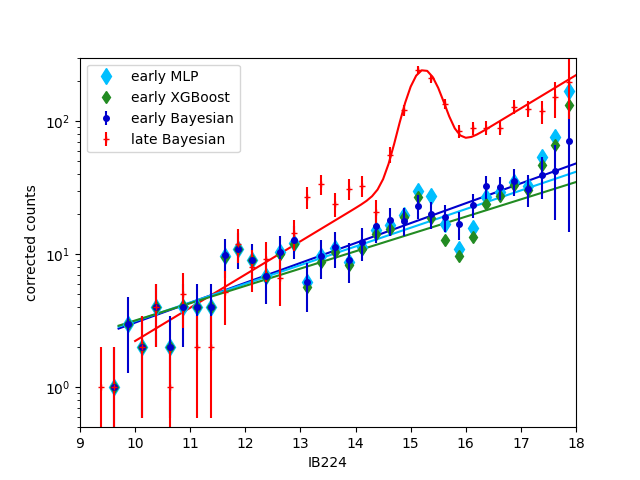} 
\includegraphics[width=\columnwidth]{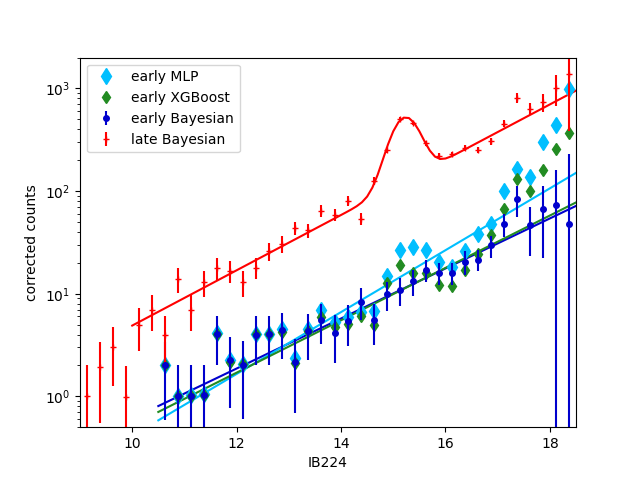} 
\caption{Completeness corrected luminosity function and fits. The dataset is divided into two regions: $R{\leq}9''$ (top) and $R>9''$ (bottom) to search for potential differences. The fits exclude the brightest bins where there is some deviation from linearity, with the count differing in the centre and further outside. Notably, for the dominating late component, the different algorithms produce very similar results.
}
  \label{Fig:klf2}
\end{figure}

We sum up the probabilities in magnitude bins (bin size = 0.25 mag) for the KLF and show them in Fig.~\ref{Fig:klf1}. The probabilities of spectroscopically classified were previously set to  1.0 (early) or 0.0 (late).  We excluded areas that were only covered by two exposures, because of their higher incompleteness. For late-type stars, the three methods generally agree. We therefore show only the late-type KLF obtained with the Bayesian method.
The agreement between the early-type KLFs is strong for early-type stars brighter than $m=15$\,mag. However, MLP and, to a lesser extent, XGBoost show a bump around 15.25, coinciding with the RC. This is not surprising, because these methods are unaware of the RC, contrary to the Bayesian approach, and therefore struggle to take it into account. The data becomes increasingly incomplete at $m\gtrsim16.5$\,mag. 
To estimate and correct completeness, we used the KLF of \citet{gallego2018distribution} and employed the inverse of the completeness as weights.
We binned the KLF of \citet{gallego2018distribution} and scaled it so that the median bin count between 11 and 16 magnitudes was equal to our total (early plus late) KLF. Up to 16.25 magnitudes, the curves are very similar, thus our completeness is close to 100\%. At fainter magnitudes, our number counts decrease and we used the differences between the KLFs for completeness correction. We now proceed to analyse the completeness-corrected KLFs (see Fig.~\ref{Fig:klf2}). 

For fitting the KLFs, we omitted the brightest bins, where the counts are small and there is a deviation from the approximate power-law used by us. We fitted the remaining star counts down to the limit $m=18.5$\,mag\footnote{It is 18.0 for within 9$''$, because of zero counts in the 18 to 18.25 bin, in practice the weight of the faintest bin is very low because of the large uncertainties.} with a power-law function for early-types and a power-law plus Gaussian-one (to account for the RC) for late-types. For the fits, we used the square root of the counts as the absolute error.

The best-fit value for the slope of the power law is $\delta\approx0.25$ for the late-type KLF with only small variations (at most 0.02) by classification method and radial range. 

\begin{table}
\centering
\caption{KLF\ fits}
\label{Tab:klf_fits} 
\begin{tabular}{llll}
\hline
\hline
Method & range& a & $\delta$  \\
\hline
MLP & all &39.6$\pm$1.6 & 0.20$\pm$0.01 \\
XGBoost & all &  31.1$\pm$1.4 & 0.17$\pm$0.01  \\
Bayesian & all &  34.5$\pm$1.5 & 0.18$\pm$0.01  \\
MLP & <9 & 18.5$\pm$1.1 & 0.14$\pm$0.01 \\
XGBoost & <9 &  16.5$\pm$1.0 & 0.13$\pm$0.01    \\
Bayesian & <9 &  20.3$\pm$1.1 & 0.15$\pm$0.01   \\
MLP & >9 & 18.8$\pm$1.1 & 0.30$\pm$0.02  \\
XGBoost & >9 &  13.3$\pm$0.9 & 0.26$\pm$0.02   \\
Bayesian & >9 &  13.3$\pm$0.9 & 0.24$\pm$0.02   \\

\hline
\end{tabular}
\tablefoot{Power-law fits of the early-type KLFs employing different radial ranges. The density $a$ is defined as the number of stars per magnitude at 15.5. $\delta$ is the power-law exponent.

}
\end{table}

The results for early stars are presented in Table~\ref{Tab:klf_fits}. When using all the bins and the full radial range, the obtained best-fit parameters are fairly consistent. Although the slope increases from XGBoost to MLP, the difference is not large.

\begin{figure} [ht!]
\centering
\includegraphics[width=\columnwidth]{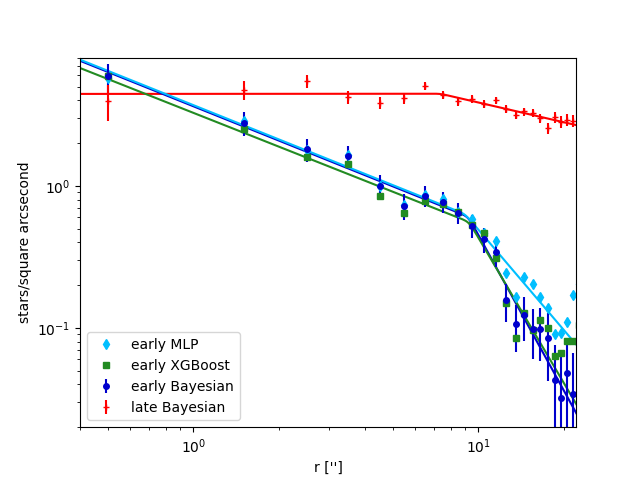} 
\caption{Radial distributions obtained with the different methods, summing up probabilities. The dominant late component shows similar results across different algorithms and exhibits an opposite trend to the early component. 
}
  \label{Fig:r1}
\end{figure}

Next, we analyzed by dividing the dataset into inner and outer areas at $R=9''$, the break radius for the early-type stars (see Fig.~\ref{Fig:r1}). The best-fit parameters remain relatively consistent concerning the different classification methods when considering only stars within $R=9''$. However, the KLF slope becomes significantly steeper outside of $R=9''$. 

For studying the radial profile, we excluded all stars fainter than 16 magnitudes and those situated in regions where only two observed frames are available for a given epoch (specifically, the four corners), as illustrated in Fig.~\ref{Fig:r1}. 

We fit the distributions with broken power laws. The best-fit parameters are provided in Table~\ref{Tab:radial_fits}. The early-type stars show a much steeper profile at all distances than the later-type stars.

\begin{table}
\centering
\caption{Radial fits}
\small
\label{Tab:radial_fits} 
\begin{tabular}{llllll}
\hline
\hline
Method & type& r(break) & dens & $\beta$ & $\gamma$  \\
\hline
MLP & late & 7.3$\pm$1.1 & 4.4$\pm$0.3 & 0.49$\pm$0.07  & 0$\pm$0.08\\
XGBoost & late & 7.0$\pm$1.0 & 4.5$\pm$0.2 & 0.47$\pm$0.06  & 0.01$\pm$0.07\\
Bayesian & late & 7.4$\pm$1.2 & 4.4$\pm$0.2 & 0.46$\pm$0.07  & 0$\pm$0.08\\

MLP & early & 9.0$\pm$1.0 & 0.6$\pm$0.1 & 2.36$\pm$0.35  & 0.79$\pm$0.08\\
XGBoost & early & 9.2$\pm$0.7 & 0.5$\pm$0.1 & 3.41$\pm$0.51  & 0.8$\pm$0.08\\
Bayesian & early & 9.2$\pm$0.6 & 0.6$\pm$0.1 & 3.65$\pm$0.47  & 0.81$\pm$0.08\\
\hline
\end{tabular}
\tablefoot{Parameters obtained in radial fits using a Nuker profile adapted to an essentially broken law, with $\alpha$ constrained to 64.
}
\end{table}

As can be seen in Fig.~\ref{Fig:r1}, at $R>10''$ the numbers of early-type candidates start to diverge between the results of the Bayesian and MLP methods. At larger $R$, XGBoost also provides somewhat larger early-type numbers than the Bayesian method. This raises doubts about the validity of the candidates found exclusively by MLP. To explore and gain a more profound understanding of this issue, we undertook an in-depth analysis of the SEDs of early-type candidates identified by the MLP method, particularly focusing on those in the outermost regions (see Appendix~\ref{SED_early}). Figure~\ref{fig:sed_young_cand_mlp} shows the SEDs of three early candidates located in the corners of the image (see Fig.~\ref{Fig:map_candidates}). The CBD values and the distinctive shapes of their SEDs suggest characteristics typical of early-type stars. Furthermore, within these specific regions, late-type candidates exhibit a pronounced dip corresponding to the CO band head in their SEDs. Therefore, we conclude that some of the stars exclusively identified by the MLP method may, indeed, represent authentic young stars.

\section{Discussion}\label{sec:discussion}

\subsection{Distribution of the stars in the Galactic Centre}\label{sec:dist_stars}

We identified 155 new early-type candidates and 4806 late-type candidates, all consistently identified by the three methods and previously lacking spectroscopic identification. While the majority of early-type candidates identified through the three methods are concentrated in the innermost region (R<$14''$), it is worth noting a significant number of candidates detected in the outer region (20 identified by the Bayesian method and 147 by the MLP method, see Fig.~\ref{Fig:map_candidates}). This outer region has been less explored in previous studies. Notably, the MLP method indicates an increased density of early-type stars at faint magnitudes and larger distances (see Table~\ref{Tab:can}). As discussed earlier, some of these candidates may indeed be true (see Appendix~\ref{SED_early}), but this requires further investigation. Therefore, focusing on stars identified as early-type by all three methods ensured reliability and avoided misidentifying early-type stars. However, it is important to note that our final list of stars presents the probabilities of being early-type stars according to each of the three methods.

\begin{figure} [ht!]
\centering
\includegraphics[height=0.9\textheight]{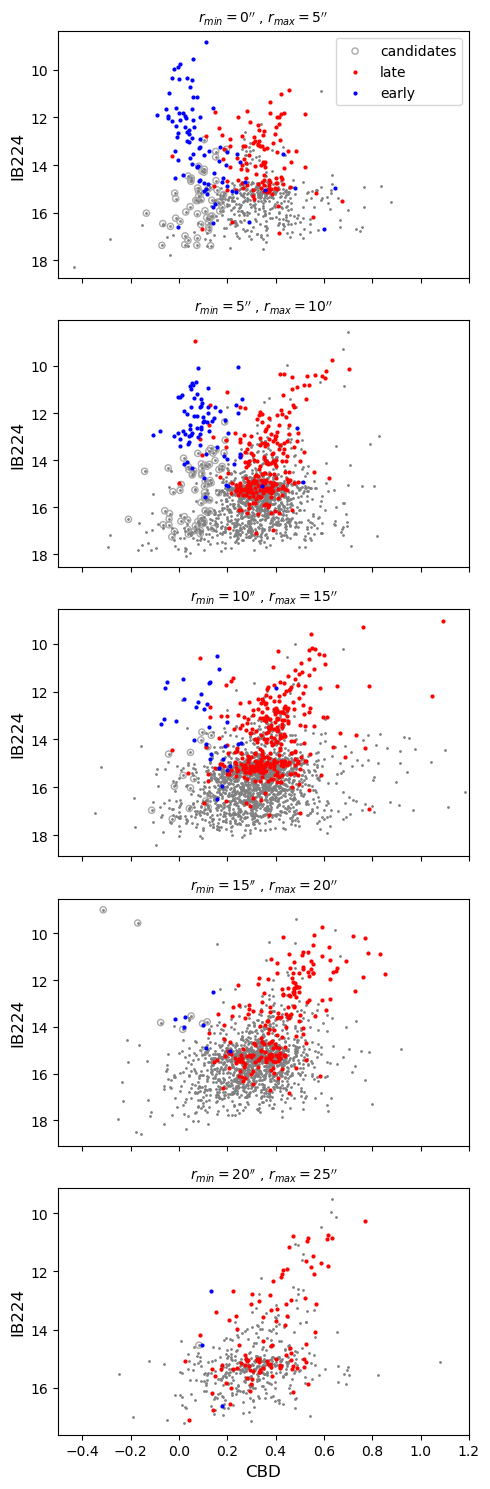} 
\caption{CBD diagrams computed for stars in different rings around Sgr\,A* highlight the diminishing influence of the young bright branch as the distance from the black hole increases.
}
  \label{Fig:cbd_rings}
\end{figure}

Figure \ref{Fig:cbd_rings} displays CBDDs within different concentric rings around Sgr A*. The illustration features identified common early candidates (grey circles), classified early-type stars (blue points), and classified late-type stars (red points). As described in the previous section, the count of early-type stars diminishes with increasing distance from the black hole.

To provide an initial assessment of the completeness in identifying the different stellar types based on the spectroscopic sample we rely on, we simulated a cluster by using the Python package \texttt{SPISEA} \citep{hosek2020spisea}. We assumed a star formation history similar to that inferred by \cite{schodel2020milky}, with 1\% of the stellar mass forming at 4 Myr, 4\% at 100 Myr, 15\% at 3 Gyr, and 80\% at 10 Gyr, we applied a metallicity value of 0.15 and constant photometric uncertainties of 0.03 mag. We converted star magnitudes to our CBD values using a linear function for the early-type stars and a second-order polynomial function for the late-type stars. Our analysis reveals a high level of completeness across varying magnitude ranges (100\% for stars <15 mag, 98\% for 15-15.5 mag, and 90\% for 15.5-16 mag).

We now compare our results to the literature. Our IB photometry results cannot be compared to \cite{buchholz2009composition} as their data was not published, and \citet{plewa2018random} only provided uncalibrated data without uncertainties. It is worth noting that our study, along with the latter, utilizes a probabilistic approach in assigning probabilities of early type to stars, unlike the former classification-based method. \citet{plewa2018random} identify 363 early-type candidates ($p>0.5$). We successfully matched 2756 of their 3165 stars with our catalogue, including 303 of their early-type candidates. The number of spectroscopically classified stars is greater in our catalogue than in the earlier work of \citet{plewa2018random} and we can therefore cross-check their results. 28 of their early-type candidates are spectroscopically classified as late-type, and 64 as early-type. As for their candidates without a spectroscopic match (211), the majority is not early according to our probabilities, with 74\% in the case of our Bayesian method. The majority of the early stars from \cite{plewa2018random} do not lie in the early-type region of our CBDD. We conclude that the differences between their study and ours are significantly larger than the differences between our 3 methods at these magnitudes. To assess whether the difference is random or caused by a systematic issue, we counted the number of early stars in four bins of magnitude and radius, as shown in Table~\ref{Tab:plewa_esum}. There are systematic differences visible, with more early stars detected, particularly noticeable for faint and outer stars, compared to our three methods, all of which show close agreement. Additionally, there is a blue clump that is more prominent than in all of our datasets. This indicates suboptimal analysis, especially for faint stars. Another reason for the difference is that \cite{plewa2018random} used only the early stars from \cite{paumard2006two}, which included very few fainter early stars, making extrapolation more important in his analysis.
We believe that the main reason for these differences lies in the way the data were reduced and analysed. We have set serious emphasis on careful photometry with robust uncertainty estimation as well as on exploring different ways of analysing the data.

\begin{table*}
\centering
\caption{Summed early probabilities of matched stars with \cite{plewa2018random}.}
\label{Tab:plewa_esum} 
\begin{tabular}{lllll}
\hline
\hline
Method &  $m<14.5$ \& $R<11''$ & $m>14.5$ \& $R<11''$ & $m<14.5$ \& $R>11''$ &$m>14.5$ \& $R>11''$ \\
\hline
Plewa & 162.1 & 139.1 & 116.3 & 141.9\\
MLP & 188.2 &  84.3 & 56.9 & 54.2 \\
XGBoost & 175.1 & 75.0 & 43.4 & 35.2  \\
Bayesian &  199.6 & 71.7 & 44.0 & 23.1  \\
\hline
\end{tabular}

 \end{table*}

\cite{buchholz2009composition} identify 312 new early-type candidates in addition to the 90 early-type stars that were spectroscopically identified at the time.  We successfully matched 3908 of their 5914 stars with our catalogue, including 280 of their early-type classifications. We cross-checked their classifications with our compilation of spectroscopic data. 19 of their early-type candidates are spectroscopically classified as late-type, 139 as early-type. As for their 124 candidates without a spectroscopic match, the majority is not early according to our probabilities, with 59\% in the case of our Bayesian method. This indicates better performance compared to \citet{plewa2018random}, aligning more closely with our findings. 
That is also confirmed when comparing the total number of early stars, as shown in Table~\ref{Tab:buchholz_esum}. In most bins, the \cite{buchholz2009composition} sum is between our 3 variants, except for faint stars further out - the most challenging region with the fewest stars. It is also possible that some of our early probabilities are unclassified by their work. Thus, this time, the main contributing factor is random differences in the CBD values. Similar to the case in \cite{plewa2018random}, the agreement is minimal for most RC stars and some brighter stars located beyond approximately $12''$. The increased depth also results in agreement for some stars fainter than the RC within about $7''$

\begin{table*}
\centering
\caption{Summed early probabilities of matched stars with \cite{buchholz2009composition}.}
\label{Tab:buchholz_esum} 
\begin{tabular}{lllll}
\hline
\hline
Method &  $m<14.5$ \& $R<11''$ & $m>14.5$ \& $R<11''$ & $m<14.5$ \& $R>11''$ &$m>14.5$ \& $R>11''$ \\
\hline
Buchholz & 141.8 & 99.6 & 33.5 & 13.0\\
MLP & 155.5 & 102.6  & 36.4 & 42.7 \\
XGBoost & 144.7 & 88.5 & 28.1 & 24.5 \\
Bayesian &  166.5 & 90.6 & 29.1 & 18.8  \\
\hline
\end{tabular}
\tablefoot{ We use a probability of 1.0 for E1, 0.98 for E2 and 0.8 for E3 for their classes aligning at their $\sigma$ definitions of the classes. Since most are E1 the precise values are not so important.}
\end{table*}

\subsection{K-band luminosity function}\label{sec:klf}

We determine a slope parameter for the KLF of late-type stars, obtaining a value of 0.25. This parameter exhibits only minor variations across different classification methods and radial ranges. Compared with previous studies, \cite{buchholz2009composition} reported a steeper slope of 0.31$\pm$0.01 using stars up to $m=14.5$\,mag, while \cite{plewa2018random} obtained a steeper slope of 0.36 using stars up to 14.5 mag. The differences in slopes can be attributed to completeness corrections or the broader magnitude range considered in the current study. Notably, in the case of \cite{plewa2018random}, their RC is rather faint, which means that the bright end tail matters more for their fit, likely contributing to the observed steep slope. That observation is supported by the steeper slope, ranging from 0.29 to 0.32, obtained when excluding stars within the RC or those with fainter magnitudes.

For early-type stars, we observe a shallower slope compared to late-type stars, with a calculated value of 0.18 using the Bayesian method. The obtained values for the inner region (R<$14''$) and the entire range are very consistent across the three methods. However, variations become more apparent for larger distances, where the MLP method yields a steeper slope. The difference in the KLF between the region of the clockwise rotating disc of young stars ($0.8''<R<12''$) and the outer area was also identified by \cite{bartko10imf}, with a flatter profile in the disk region. 
\cite{buchholz2009composition} report a slope of $0.13\pm0.02$ for the central $7''$ and $0.14$ overall, flatter than what we find here. A possible explanation for this difference could be the limit of $m=15.5$\,mag used by \citet{buchholz2009composition}, while we fit fainter stars as we explained previously. Our KLF inside of $R=14''$ looks somewhat flatter for bright stars and steeper for fainter stars, which may explain some of the difference. In our analysis, the KLFs of early-type stars show a consistent increase with magnitude, lacking evidence for the maximum at m$_{K}{\sim}13$ reported by \cite{bartko10imf} between 0.8 and 12 arcseconds. 

\subsection{Density profiles of the stars}\label{sec:density_profile}

Our results shown in Table\,\ref{Tab:radial_fits} agree with other work, including \cite{buchholz2009composition}, \cite{bartko10imf},\cite{plewa2018random},  and \cite{do2009nocusp}. We observe a very shallow, nearly constant inner profile for late-type stars and a steeper one further out. The early type stars show a broken power law, too, but their density increases steeply towards Sgr\,A* at all distances (see Fig.\ref{Fig:r1}).

Quantitatively, the slopes reported by \cite{buchholz2009composition} for early-type stars are 1.08 $\pm$ 0.12 between $1''$ and $10''$, and 3.46 $\pm$ 0.58 between $10''$ and $20''$. However, differences become more apparent for outer early-type stars at large projected distances. Our MLP classification method -- and to a lesser degree XGBoost -- finds that the number density of early-type stars may increase again at $R\gtrsim 20"$. This very interesting possibility should be followed up with photometric and spectroscopic work.

\subsection{Temperature and metallicity of late-type stars}\label{sec:metallicity}

Both temperature and metallicity will affect the strength of the CO bandhead absorption. This raises the possibility of applying IB photometry to measure these stellar properties.

We began our analysis by examining the effective temperature of late-type stars, which is more directly related to the CBD than metallicity.

Our analysis encompasses two primary objectives: firstly, to explore the visibility and significance of warmer-than-average late-type stars on the classification probabilities, and secondly, to explore the use of CBD and CBDD for estimating temperatures and metallicities.

We excluded stars with a metallicity $[M/H]<-2$, which we do not consider reliable. Our analysis of 25 stars with data from both works revealed no significant temperature offset between them. In total, we analyzed 394 stars, 51 from \cite{do15metallicy} and the majority (318 stars) from \cite{feldmeier2017kmos}.

\begin{figure} [ht!]
\centering
\includegraphics[width=\columnwidth]{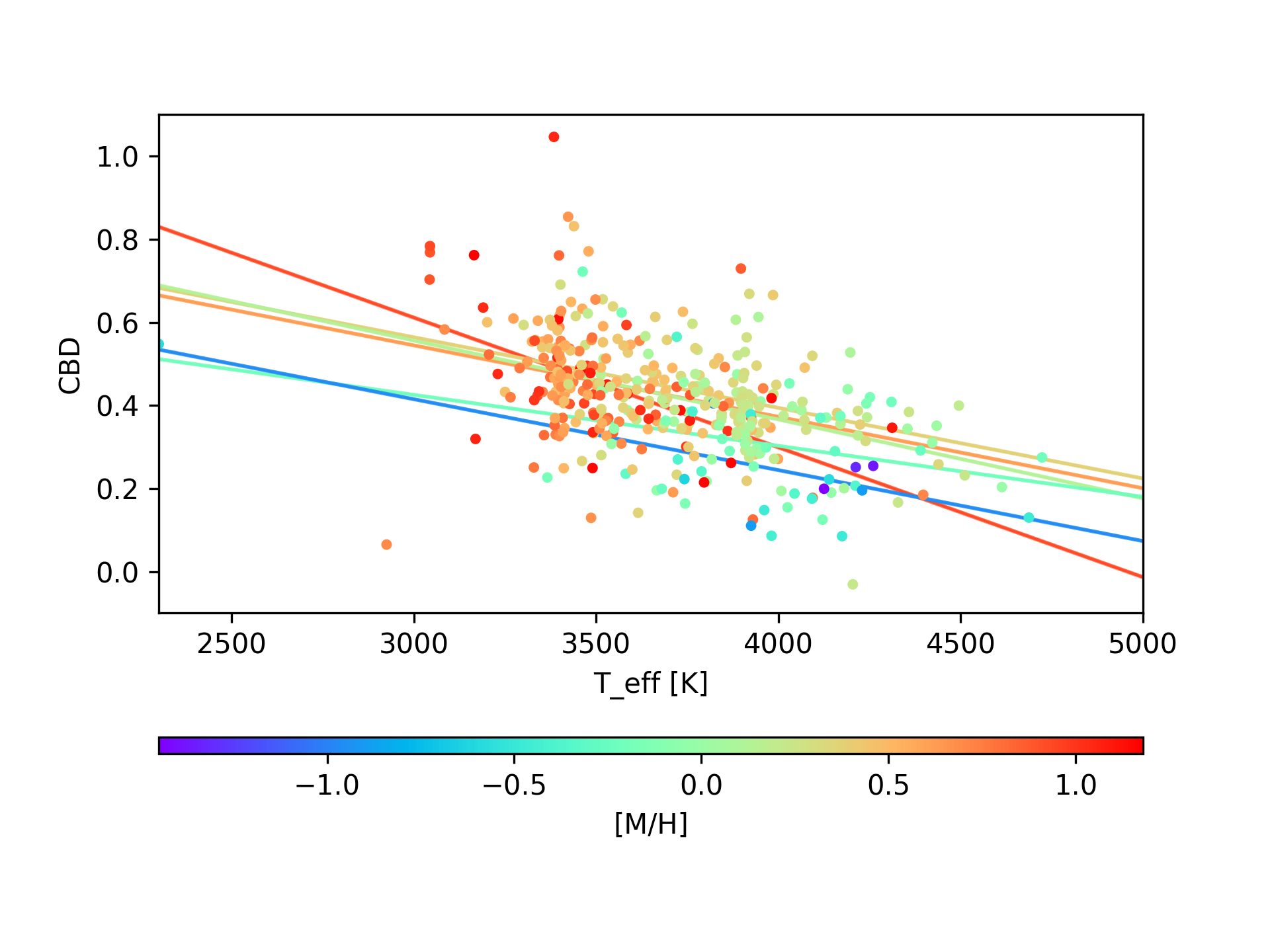} 
\caption{Relationship between the CBD value and spectroscopic effective temperature of late stars using data from \cite{do15metallicy} and \cite{feldmeier2017kmos}. Linear relationships were fitted within magnitude bins, with results depicted by the lines. The bins for the fits were categorized as follows: [M/H] < -0.5 dex (blue), -0.5 < [M/H] < 0, 0 < [M/H] < 0.25, 0.25 < [M/H] < 0.5, 0.5 < [M/H] < 0.75, and [M/H] > 0.75 (red). The colour of the lines indicates the respective mean metallicity.
}
  \label{Fig:temp_CBD}
\end{figure}

When plotting the temperatures against CBD (Fig.\,\ref{Fig:temp_CBD}) we found a linear relationship. This allows us to connect these two parameters via the following relationship:
\begin{equation}
    T_{\text{eff}} = (4157 \pm 44) + (-1109 \pm 101) \cdot \text{CBD},
    \label{eqn:T_CBD}
\end{equation}
similar to the findings in \cite{feldmeier2017kmos}. Unlike \cite{pfuhl2011SFH}, we found that higher-order parameters are unnecessary. However, it is important to note that our dataset covers a narrower temperature range and uses a different CBD definition than their work.

When we split the sample by metallicity, noticeable differences become visible, as depicted in Fig.~\ref{Fig:temp_CBD}. Between [M/H]= 0.75 and 0 dex, the fits exhibit a high degree of similarity\footnote{The discrepancy in the fit for [M/H]$>0.75$  may be attributed to the lower reliability of the spectroscopic and CBD data for these highly metal-rich stars, possibly due to the higher abundance of AGB stars in that range.}. The CBD values are shifted to lower values for lower metallicities. This is plausible, as fewer metals should result in less pronounced spectral lines. Hence, the data indicate that the mean CBD decreases as a function of metallicity. 

The dependency of CO depth on metallicity is also observed in the more narrowly defined spectroscopic CO index \citep[see][]{fritz21kmos}. In our case, the dependence is more influenced by continuum extrapolation because we measure the depth at 2.36 $\mu$m. The dependency on metallicity suggests that the two-parameter conversion of effective temperature to CBD (Equation~\ref{eqn:T_CBD}) is overly simplistic, necessitating additional parameters. We used following equation which uses also as [M/H] and [M/H] squared for convert  T$_\mathrm{eff}$ to our CBD:

\begin{equation}
\begin{split}
    CBD = & 1.08167498 -0.00018242 \cdot T_{\text{eff}} \\
    & +0.07098599 \cdot \text{[M/H]} -0.05281159 \cdot \text{[M/H]}^2.\label{eqn:T_CBD2}
\end{split}
\end{equation}

The inclusion of [M/H] squared accounts for saturation effects at high metallicity levels.

Next, we used the obtained relationship in Equation\,\ref{eqn:T_CBD2} to look into the metallicity. From the same 25 stars with metallicity as before, we found a consistent offset of 0.34 dex between the metallicities estimated by \cite{do15metallicy} and \cite{feldmeier2017kmos}. To align both datasets on the same stars, we adjusted by adding or subtracting half of this offset.

\begin{figure} [ht!]
\centering
\includegraphics[width=\columnwidth]{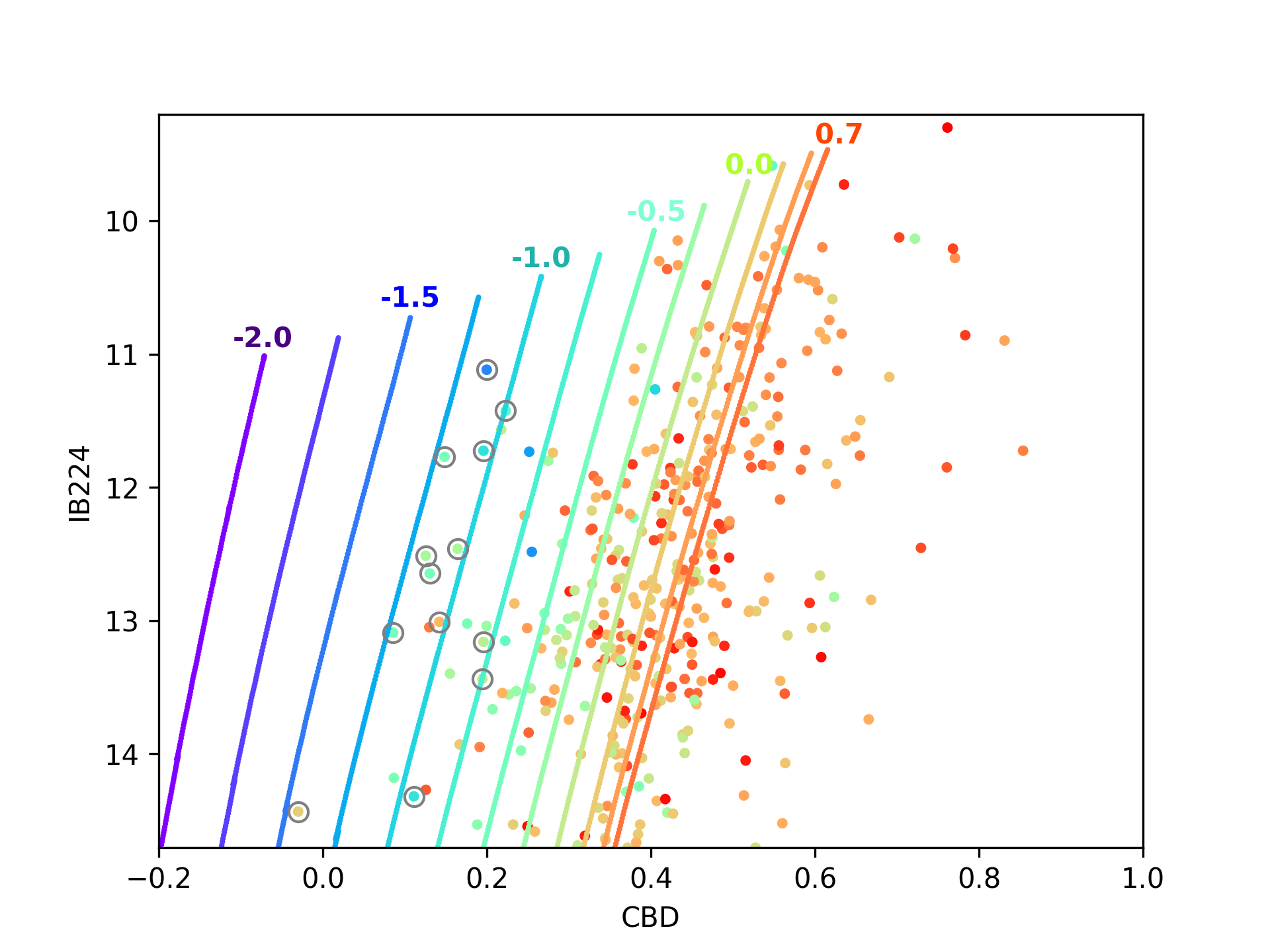} 
\includegraphics[width=\columnwidth]{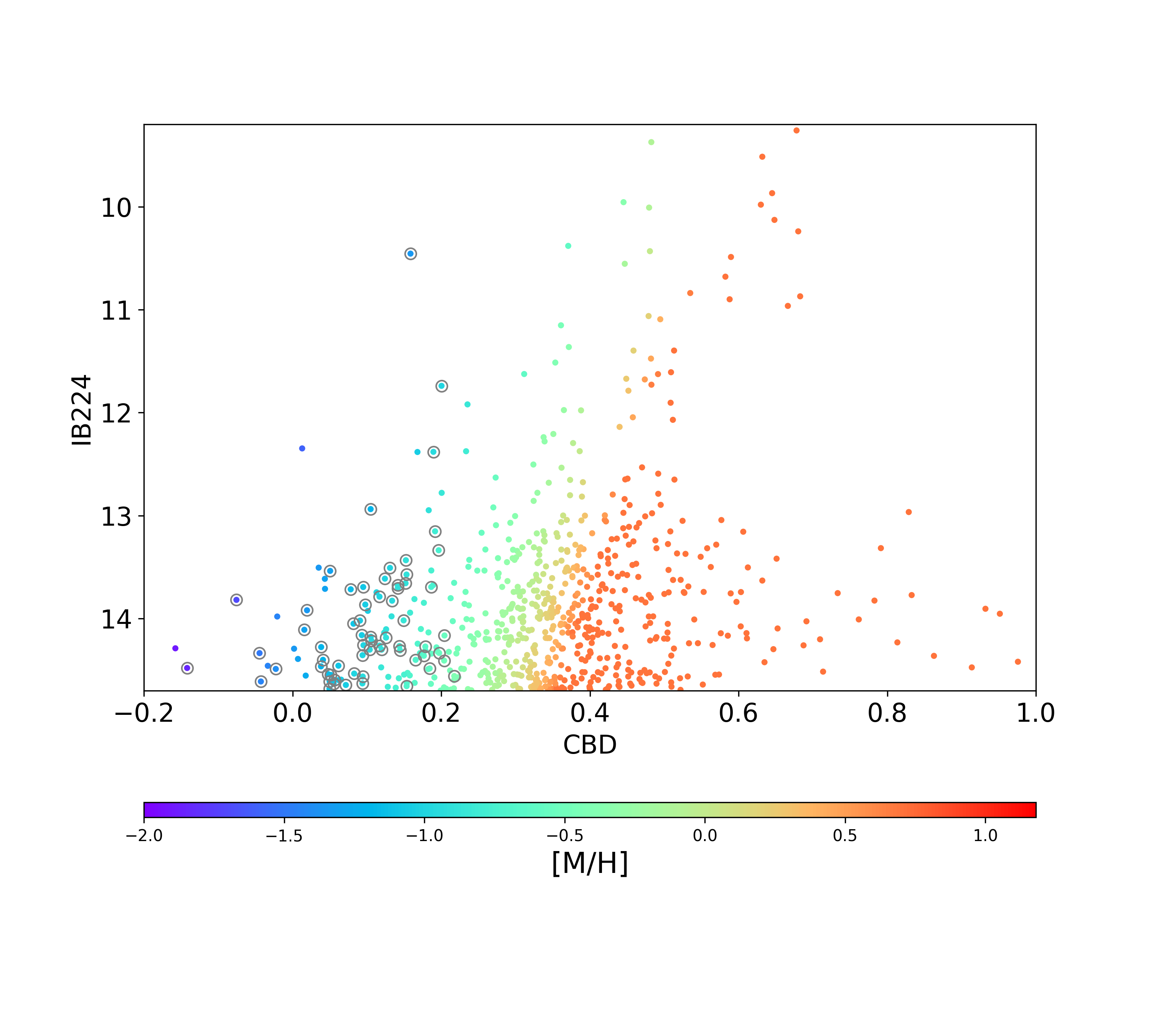} 
\caption{Metallicity estimation analysis. The top panel illustrates the CBDD of late stars with spectroscopic metallicity from \cite{do15metallicy} and \cite{feldmeier2017kmos}. 
We used 12 Gyrs Giant branch isochrones with different metallicities from \cite{bressan12parsec,chen2015parsec,marigo17parsec} to convert T$_\mathrm{eff}$ into CBD values. The bottom panel shows the CBDD diagram of stars lacking spectroscopic metallicities and not classified as spectroscopic early. Metallicities are assigned using these isochrones. Both plots utilize the same colour scale and are restricted to stars brighter than 14.7, as this is where the presence of the RC, not included in our model, becomes significant. Stars encircled in grey represent those classified as likely early by the Bayesian method.
}
  \label{Fig:met_CBD}
\end{figure}

We present the spectroscopically estimated metallicities in the CBDD shown in the top panel of Fig.~\ref{Fig:met_CBD} and overplotted 12 Gyrs\footnote{Due to the weak dependence of the giant branch on age, the precise age used in this analysis is not crucial.} isochrones from \cite{bressan12parsec,chen2015parsec,marigo17parsec}.
We focus only on the giant branch, which is the least populated region of our CBDD. The majority of stars have a metallicity $[M/H]\approx0.70$\,dex. We exclude stars fainter than 14.7\,mag, because those stars are in their majority RC stars, which are more challenging to model. We translate their spectroscopically estimated effective temperatures (T$_\mathrm{eff}$) into CBD values using Equation~\ref{eqn:T_CBD2}.

As becomes apparent from the top panel of Fig.\,\ref{Fig:met_CBD} a combination of the CBD with magnitudes offers a promising approach for estimating metallicities. We used isochrones in steps of 0.05 dex to estimate metallicities by identifying the closest neighbour on the CBDD. This resulted in a predicted [M/H] range of -1.5 to 0.7, which closely aligns with the lowest measured [M/H] of -1.44. However, the difference is more pronounced on the metal-rich side. It remains uncertain whether the very metal-rich stars are genuine, as higher spectral resolution observations \citep[e.g.][]{rich2017metallicity,do2018super,joennson20apogee,thorsbro2020detailed} do not detect them. It is evident that the results are less precise on the metal-rich side due to the closer spacing of isochrones. 

Subsequently, we applied this technique to estimate the metallicities of stars lacking spectroscopic data. We omitted stars identified spectroscopically explicitly as early-type, but retained all early-type candidates. Additionally, we excluded stars brighter than $0.5$\,mag than the top of the isochrones because these could be AGB stars or may have unreliable photometry. 

We thus predict the metallicities of 700 stars. Among them, 120 are classified as metal-poor with [M/H] <-0.5, with approximately half of them identified as preferable early. When excluding preferable early stars, as identified through the Bayesian method, 58 metal-poor stars remain. Further exclusion of preferably early-type stars identified through any of our classification methods results in a sample of 617 stars. In this latter case, 6.3\% of the likely late-type stars are classified as metal-poor.

The fraction of low-metallicity stars determined from our photometric data aligns with spectroscopic assessments suggesting that warm,low-metallicity stars comprise up to 10\% of the total population, as reported in studies such as \cite{pfuhl2011SFH,do15metallicy,feldmeier2017kmos}. While our estimate is in line with these findings, it remains uncertain due to the potential inclusion of early-type stars in our sample, as well as the likelihood of overlooking low-metallicity late-type stars, which may be misclassified as early-type due to weaker CO band heads, see also Appendix~\ref{app:metallicity}. Addressing both factors comprehensively would necessitate modelling the spatial and magnitude distributions of all three classes (late metal-rich, late metal-poor, early), a task that falls beyond the scope of this paper.

\subsection{IMF of early-type stars}

To infer the IMF of the young cluster formed with the early-type candidates brighter than 16 magnitudes and located within a distance smaller than the computed break radius of 9$''$, we employed an optimization algorithm that compares observational data with theoretical simulations. Utilizing \texttt{SPISEA} \citep{hosek2020spisea}, we constructed a grid of synthetic single-age star clusters varying in age, cluster mass, extinction, and IMF slope index. To generate synthetic photometry for each star, we utilized the NACO K$_{S}$ filter, assuming a distance of 8.25 kpc and a metallicity of [M/H] = 0.15. SPISEA allows us to consider multiplicity in the generated clusters by constructing a multiplicity object (`multiplicity.MultiplicityUnresolved') based on \cite{lu2013stellar}. This configuration treats star systems as unresolved, combining all components into a single star within the cluster. Due to the absence of stars with lower masses in our observed magnitude ranges, the mass limits for stars in the simulated clusters were set to 5 M\textsubscript{\(\odot\)}-120 M\textsubscript{\(\odot\)}.
We used the MIST \citep[MESA Isochrones \& Stellar Tracks;][]{choi2016mesa} evolution model and adopted the reddening law from  \cite{nishiyama2009interstellar}. We examined isochrones spanning a range of cluster ages from 2 to 8 Myr and constructed synthetic KLFs for each combination of parameters. Subsequently, we compared the theoretical KLFs to the observed KLF by computing the chi-squared statistic. Finally, we identified the parameter set that minimizes the chi-squared value. The optimal fit corresponds to a cluster age of 4 Myr. We utilized a Monte Carlo (MC) approach to incorporate the uncertainty arising from the sampling of the IMF in each realization of a cluster with SPISEA. It is important to note that the choice of the evolution model can impact the fit values, particularly the inferred age.
The best power-law IMF index value is $1.6 \pm 0.1$, obtained for the three sets of early-type candidates using different methods. To validate our results, we also fitted the IMF by limiting the magnitude range up to 15. In this analysis, we determined that the IMF slope is approximately 1.7, falling within the expected range and accounting for uncertainties in our results.

\begin{figure} [ht!]
\centering
\includegraphics[width=\columnwidth]{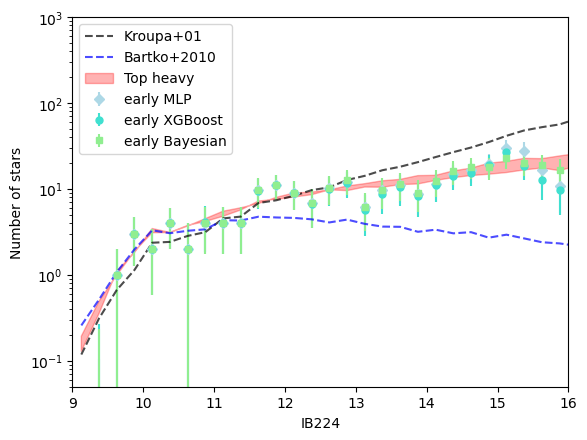} 
\caption{Comparison between different models simulated with \texttt{SPISEA} and the observed KLF for the different methods summing up probabilities and corrected by completeness (see Section\,\ref{sec:results_structure}). The red line represents the KLF corresponding to a simulated cluster with the optimal power-law IMF, featuring an IMF slope of -1.6, while the shaded region indicates the uncertainty in the model. Additionally, the KLFs of cluster models computed with SPISEA using the Kroupa IMF \citep{kroupa2001variation} and employing a top-heavy IMF \citep{bartko10imf} are presented for comparison.}

  \label{Fig:simul_imf.png}
\end{figure}

Figure \ref{Fig:simul_imf.png} illustrates the constructed KLF derived from our early-type candidates utilizing different methods. The red shaded region indicates both the optimal model fit and its associated uncertainty. Additionally, we present cluster models computed with \texttt{SPISEA}, employing the IMF slopes derived from \cite{kroupa2001variation} and \cite{bartko10imf} for comparison.

We also conducted an analysis of the IMF for distances greater than 9$''$. In this scenario, notable differences emerged in the best-fit models obtained through the three different methods, consistent with our expectations outlined in Section\,\ref{sec:klf}. The IMF slopes for Bayesian, RF, and MLP methods were 1.8, 2.0, and 1.9, respectively, suggesting larger values for regions beyond the inner regions. The optimal fit corresponds to a cluster age larger than 5.6 Myr for the three methods. This disparity suggests distinct origins for stars in the inner and outer regions, implying diverse mechanisms driving star formation across the NSC. The deviations from the standard Salpeter/Kroupa IMF (slope -2.3) observed in the central region suggest unique formation processes, while the potential presence of a standard IMF further out indicates an evolving environment. These findings emphasize the complexity of star formation dynamics within the cluster and highlight the influence of local conditions on stellar birth.

\section{Conclusions}\label{sec:conclusions}
In this paper, we classify stars within the central parsec around Sgr~A*, utilizing NIR IB images obtained with NACO/VLT. While these data have been previously investigated, we revisit their analysis, improving the reduction and photometric processes, leading to enhanced sensitivity and more reliable estimates of photometric uncertainties compared to prior efforts. Additionally, we employ three distinct yet complementary classification methods for distinguishing stars as early or late type: Bayesian inference, a basic neural network, and a fast gradient-boosted trees algorithm. Thanks to the information from extensive spectroscopic studies conducted over the past decade, we expand our database, enhancing the training dataset for our models and thereby obtaining more robust and confident results.

Despite challenges associated with studying this region, such as extreme interstellar extinction and crowding, we identify 155 new early star candidates using Bayesian and MLP methods. The identified candidates present promising targets for upcoming spectroscopic observations, which are crucial for validating their classification and providing additional support for our results. These candidates, while intriguing as potential spectroscopic targets, are not the primary outcome of our probabilistic approaches. 

Consistent with previous work we find that broken power laws provide adequate fits to the radial density distribution of the late and early-type stars. Our investigation reveals a break radius of $7.4\pm 1.2''$ for the late-type distribution and $9.2\pm 0.6''$ for the early-type distribution. The slopes for the late-type distribution are determined to be $\beta =0.46\pm 0.07$ (outside of the break radius) and $\gamma = 0\pm 0.08$, while the early-type distribution exhibits slopes of $\beta =3.65\pm 0.47$ and $\gamma = 0.81\pm 0.08$.

The comparison with spectroscopic temperature reveals that our CBD primarily follows a linear trend with temperature, although metallicity also plays a significant role for [M/H]<0. When predicting metallicities for over 600 stars, we find that approximately 6\% exhibit metal-poor characteristics ([M/H]<-0.5), a result consistent with spectroscopic metallicities.

We find that the mass function of the early type stars appears to be top-heavy near Sgr\,A*, with a power-law slope of $1.6 \pm 0.1$. Contrary to \citet{bartko10imf} we find no peak of the luminosity function around 14\,mag and a significantly steeper mass function. This is probably related to the incompleteness of the early spectroscopic data at faint magnitudes.

At larger projected distances, the best-fit models vary significantly among the methods, with all IMF slopes exceeding 1.6 and possibly in agreement with a standard Salpeter/Kroupa IMF. These findings suggest distinct star formation mechanisms between the inner and outer regions, with deviations from the standard Salpeter/Kroupa IMF indicating unique processes in the central region and a potentially evolving environment further out.

The observed slope and profile of early-type stars are consistent with the formation of the young stars near Sgr\,A* on a gaseous disc \citep{Bonnell:2008ys}. At larger distances, other formation mechanisms may have been at work.

We demonstrate the versatility of IB photometry, which not only aids in identifying promising targets for upcoming spectroscopic observations but also plays a crucial role in validating their classification. Moreover, IB photometry enables the exploration of important stellar features such as metallicity and temperature, providing valuable insights into the formation history of the NSC. Consequently, the importance of IB photometry extends to the design of IB filters for telescopes.

\begin{acknowledgements}
This work is based on observations made with ESO Telescopes at the Paranal Observatory under programs 073.B-0084(A), 073.B-0745(A), 077.B-0014(A). EGC and RS acknowledge financial support from the Severo Ochoa grant CEX2021-001131-S funded by MCIN/AEI/ 10.13039/501100011033, from grant EUR2022-134031 funded by MCIN/AEI/10.13039/501100011033 and by the European Union NextGenerationEU/PRT, and from grant PID2022-136640NB-C21 funded by MCIN/AEI 10.13039/501100011033 and by the European Union. AFK acknowledges funding from the Austrian Science Fund (FWF) [grant DOI 10.55776/ESP542].
\end{acknowledgements}

%
%

\bibliographystyle{aa} 
\bibliography{bibliography.bib} 

\begin{appendix}

\section{Uncertainty estimation}\label{unc_estimation}
As seen previously, obtaining robust and reliable uncertainties without loss of sensitivity is crucial for fulfilling the objectives of our work. We developed a dedicated procedure to compute the uncertainties by using the bootstrapping method \citep{efron1979bootstrap,andrae2010error}. We applied two different approaches depending on the epochs. On the one hand, for epochs with a large (N greater than 96 exposures) number of observed frames (see in Table\,\ref{Tab:Obs}), we applied the normal procedure, as is described in \cite{gallego2022accurate}. On the other hand, we developed a new {\it noise} bootstrapping procedure for epochs where the number of exposures is eight.

\subsection{Bootstrapping}\label{boots}
\begin{figure} [ht!]

\includegraphics[width=\columnwidth]{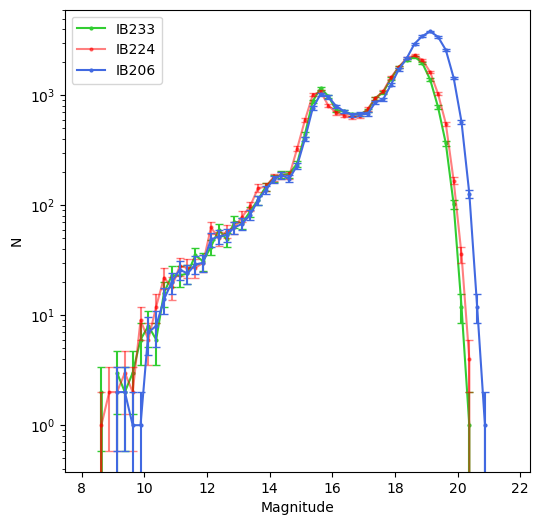}
\caption{KLFs for the final bootstrapping mosaics. The different colours correspond to the different bands.}
    \label{Fig:klf_normalboots}  
\end{figure}

\begin{figure*}[!ht]

\centering
\includegraphics[width=\textwidth]{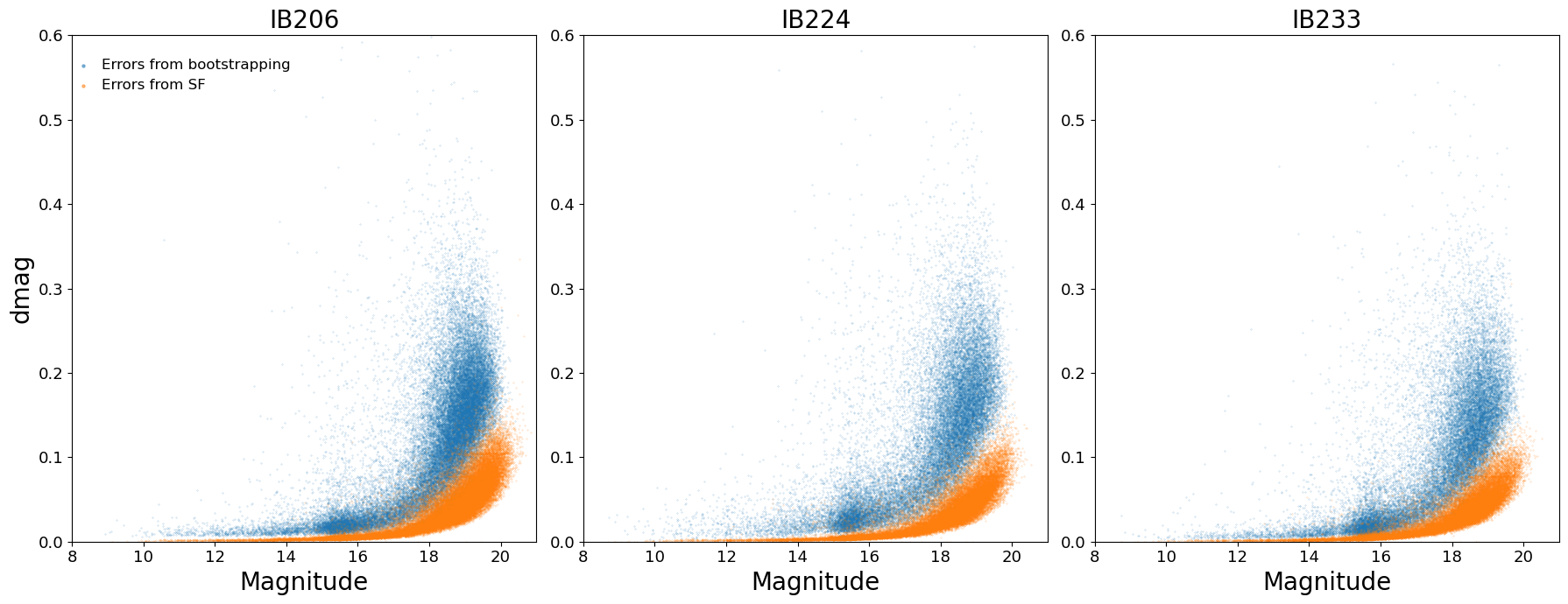}
\caption{Comparison between photometric uncertainties versus magnitude obtained with the bootstrapping procedure (blue points) and {\it StarFinder} program (orange points) for IB206, IB224, and IB233.}
    \label{Fig:comparison_bootsf}
\end{figure*}

For IB206, IB224, and IB233 epochs, we created $100$ mosaic images using the bootstrapping with replacement method. The number of frames considered in each bootstrap image is equal to the total number of frames observed for the epoch. Afterwards, we repaired the saturated stars as described in  Section\,\ref{sec:saturated_stars}, and detected and subtracted the point sources from $100$ mosaic images of each epoch following the photometric analysis described in Section\, \ref{sec:psf_fitting}. We defined the {\it detection frequency} parameter as the percentage of bootstrap deep images where each star is detected and we obtained a final list with all the stars detected in all the images and their associated detection frequency value. We selected {\it detection frequency}=50\%, which means that we considered stars detected in 50\% or more of the mosaic deep images. Finally, we computed uncertainties from the standard deviation of the position and flux of each star. Figure\,\ref{Fig:klf_normalboots} shows a comparison between the KLFs for the three epochs after the bootstrapping procedure. In Fig.\,\ref{Fig:comparison_bootsf} we show a comparison between the photometric uncertainties obtained with this method and the error from {\it StarFinder} program. In the figure, we can see that {\it StarFinder} tends to underestimate uncertainties, whereas our approach enables us to obtain reliable uncertainty estimates without compromising sensitivity.

\FloatBarrier

\subsection{Bootstrapping with noise}\label{noise_boots}
\begin{figure} [ht!]

\includegraphics[width=\columnwidth]{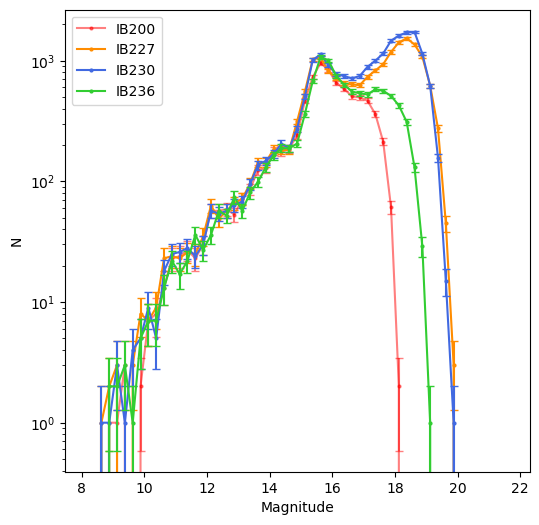}
\caption{KLFs for the final {\it noise} bootstrapping mosaics. The different colours correspond to the different bands.}
    \label{Fig:klf_noiseboots}
\end{figure}

\begin{figure*}[!ht]
\centering
\includegraphics[width=\textwidth]{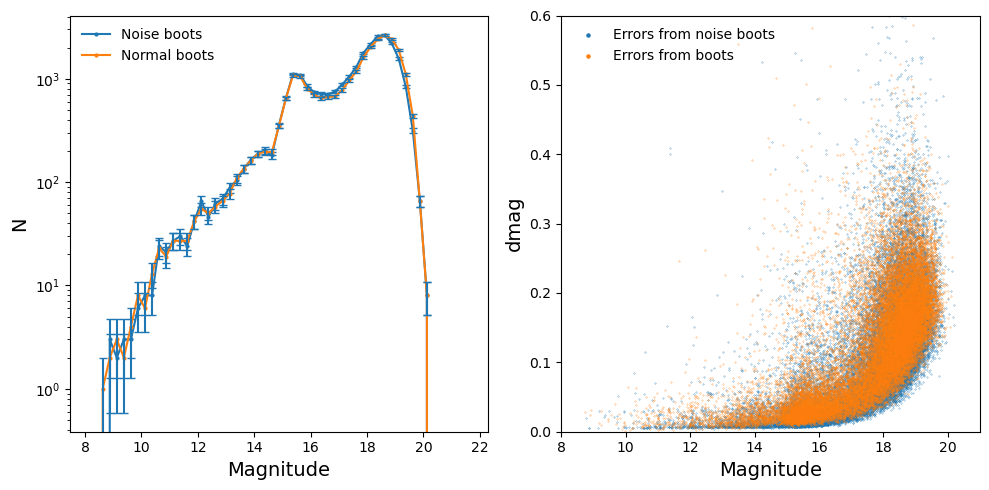}
\caption{Comparison between the results obtained by using the normal bootstrapping procedure and the new bootstrapping for IB224. {\it Left:} KLF. {\it Right:} Photometric errors.}
    \label{Fig:comparison_boot_noisebootIB224}
\end{figure*}

\begin{figure*}[!ht]
\centering
\includegraphics[width=\textwidth]{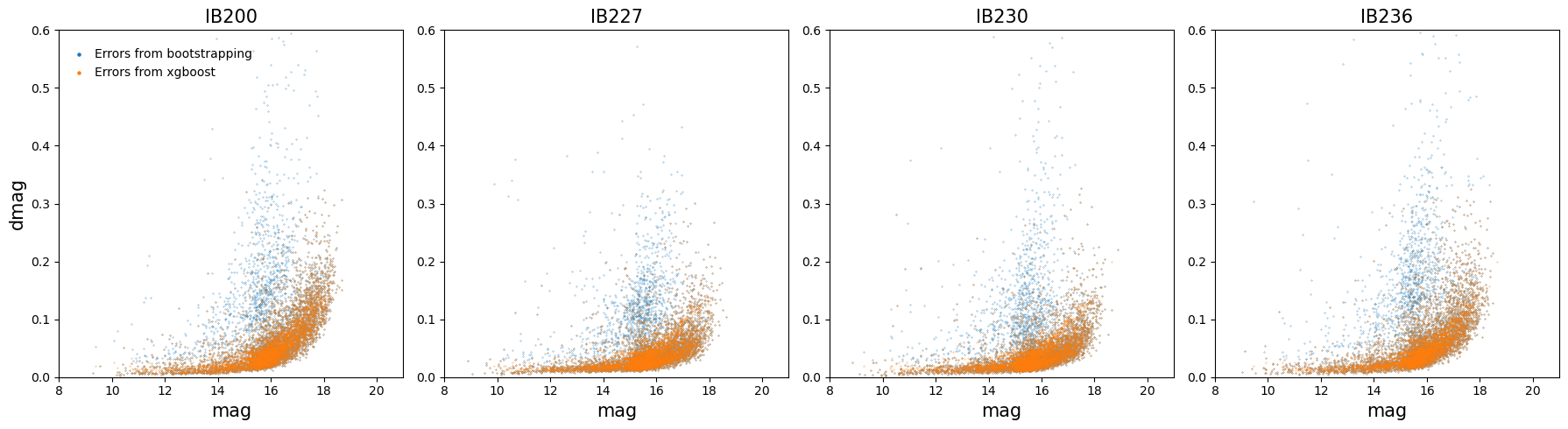}
\caption{Comparison between photometric uncertainties versus magnitude obtained with the bootstrapping  (blue points) and \texttt{XGBoost} (orange points) procedures, respectively, for IB200, IB227, IB230, and IB236.}
    \label{Fig:comparison_noisexgboot}

\end{figure*}

For IB200, IB227, IB230, and IB236 epochs, we cannot apply the previous method because the number of frames is not large enough. Therefore, we created $100$ mosaic images by combining random noise attributed to both the photon noise from the signal and the read noise from the detector. In the first place, to assess the validity of the approach, we applied to IB224 whose uncertainties were computed using the normal procedure and compared the results. Figure\,\ref{Fig:comparison_boot_noisebootIB224} shows a comparison between the KLFs and the photometric errors, respectively, obtained with the two different approaches. We obtained very similar results with the new procedure. We applied this method for the rest of the epochs to obtain the errors. Figure\,\ref{Fig:klf_noiseboots} shows a comparison between the KLFs for the four epochs after the new procedure. Figure \ref{Fig:comparison_noisexgboot} illustrates the photometric uncertainties obtained through this method, shown as blue points. While this method allows for the estimation of realistic photometric uncertainties, a distinct pattern becomes apparent in the four plots. This pattern is observed for stars with brightness between $15$ and $16$ magnitudes, where errors exceed $0.1$, and are situated at the four corners of the final mosaics, where the count of observed frames is limited to 2.

\FloatBarrier

It is important to note that this pattern is not a reflection of realistic conditions; rather, it is linked to the fact that the mosaic in those regions contains only twice as many frames as the individual images. This results in the generated images having excessive noise. Excessive errors in certain bands are undesirable because they lead to the practical effect of predominantly relying on bands with smaller errors, often limited to only three bands. As a result, the obtained values become more uncertain. To improve error predictions, we trained a machine learning algorithm, \texttt{XGBoost}\footnote{\url{https://xgboost.readthedocs.io/}} library, as detailed in \cite{chen2016xgboost}, utilizing our dataset. The target feature for prediction was the error in each of the four problematic bands (IB200, IB227, IB230, IB236), and each was addressed separately. We applied a logarithmic transformation to these errors because, without it, the algorithm tends to overemphasize a few extreme cases, leading to poor overall performance. The features utilized for training consisted of the magnitudes and the errors in the three other bands (IB206, IB224, IB233) with a large number of observed frames (see Table\,\ref{Tab:Obs}).

During the algorithm training, we excluded data points associated with bands having three or fewer observed frames available. This selection is based on our observation that stars with exactly three images have notably increased statistical significance compared to other stars. However, their overall influence remains limited due to the smaller number of stars with exactly three observed frames. To prevent overfitting, we employed an 80\% training set and a 20\% testing set. Primarily, we employed a nonrandom test approach by reserving the 20\% of the data that is farthest from the centre in either the RA or Dec direction. This approach establishes spatial connectivity in the corner regions, preventing the algorithm from achieving superior performance through interpolation across the gaps created by missing stars distributed throughout the image. In practice, we observed that this did not happen, and the results obtained using random and spatially connected test sets were similar. We performed grid fitting with various reg\_alpha\footnote{This parameter defines the size of subsets of the data which have the same constant results by binning. A larger reg\_alpha implies that more neighbouring data points share the same results.} values to prevent overfitting and select the one that yields the smallest standard deviation in the test set. It is crucial to incorporate both the magnitude in the target band and an estimation of the error in the good bands for accurate results. The significance of including magnitude as a feature is obvious, and the inclusion of errors in other bands is also not surprising. Many characteristics are similar or identical across different bands, such as the presence of close neighbours, the ratio of the number of images to the maximum possible number of images, and the positioning of AO guide stars. We achieved remarkably similar results whether we used the magnitude errors for all three bands or their mean. In the end, we considered the mean of the three because our dataset is not extensive enough to conclusively establish that a purely performance-based selection is superior. Therefore, we preferred the simpler and more interpretable model.

In addition to addressing stars with limited exposures, we also made corrections for stars whose errors exceed $0.33$. Such a substantial error is not feasible for 3-sigma detections, and all our sources meet this criterion due to the {\it StarFinder} settings. These significant errors are likely spurious, as they do not consistently appear across different bands. We extended this correction to the three other bands as well. For these bands, we excluded the target band from the mean error calculation. Only a small number of stars, approximately a dozen, display such substantial errors in each band. While these corrections may not be as robust as bootstrapping errors derived from high-quality data, we also mark the corrected magnitude errors with flags for future reference. In summary, we utilize the uncertainties derived from the XGBoost procedure for stars in regions where the number of frames in a given epoch is less than or equal to three. For stars located in the remaining regions, we employ the noise bootstrapping procedure to compute the uncertainties.

\subsection{Uncertainties of CBD}\label{CBD_errors}

Although we primarily relied on the \texttt{SciPy} \texttt{curve\_fit} function to estimate uncertainties in the CBD values, we also explored alternative methods, including those based on the $\chi^2$ for linear, polynomial, and exponential fits, respectively, similar to the approach outlined in \cite{gillessen2009orbits}.

The reduced $\chi^2$ typically exceeds the expected value of 1, and this difference is particularly significant for bright stars, given their smaller absolute errors. This behaviour is not unexpected, and it can be attributed to various factors, such as the complexity of star spectra compared to our simplified models or potential variability between the different images. Note that there is nearly a month between the acquisition of the first and last image, which may contribute to these discrepancies.

Firstly, we performed a fit of the median $\log(\chi^2)$ as a function of magnitude using a linear model. The predicted value obtained from this model serves as the default $\chi_a^2$. Secondly, we divided the original $\chi^2$ of each star by $\chi_a^2$ and determined the threshold above which the number of values exceeded twice the expected from the $\chi^2$ distribution for the degrees of freedom (d.o.f) of the model fit. In cases where this condition is met, the original $\chi^2$ is employed in subsequent calculations; otherwise, $\chi_a^2$ is used. Thirdly, the errors of the linear and other fits are adjusted by multiplying them with $\sqrt{\chi^2/d.o.f}$, using the previously derived $\chi^2$ value for each star. Finally, we calculated the CBD error by taking the square root of the sum of the two fit errors.

\section{Calibrators}\label{appendix_calib}
In this section, we provide the final list of 28 OB stars (see Table\,\ref{Tab:calibrators}) used in the basic calibration process, as discussed in Section\,\ref{sec:basic_calibration}. Figure\,\ref{Fig:map_calibrators} illustrates the spatial distribution of these stars across the FoV. The data in the table is sourced from \cite{feldmeier2015kmos}, except the $K_{S}$ magnitudes, which are obtained from \cite{schodel2010accurate}.

\begin{table*}[!htb]
\centering
\caption{OB stars used for the basic calibration.}
\label{Tab:calibrators}
\begin{tabular}{lrrrrrll}

\hline
ID  & RA & Dec & {$\Delta$}RA & {$\Delta$}Dec & $Ks$ & Name & Type\\
 & [$^\circ$] & [$^\circ$] & [$''$]  & [$''$]  & [mag]  & &  \\

\hline
00109 & 266.41724 & -29.008343 & -1.060 &  -1.916 & 10.65 & MPE+1.0-7.4(16S) &    B0.5-1 \\
00096 & 266.41437 & -29.007425 &  6.613 &   1.387 & 10.66 &                  &           \\
00166 & 266.41397 & -29.009418 &  7.660 &  -5.788 & 10.96 &                  &           \\
00209 & 266.41571 & -29.009912 &  3.014 &  -7.567 & 11.02 &                  &           \\
00230 & 266.41681 & -29.005444 &  0.082 &   8.521 & 11.07 &                  &      O9-B \\
00205 & 266.41882 & -29.007736 & -5.305 &   0.268 & 11.14 &            IRS1E &      B1-3 \\
00273 & 266.41684 & -29.004976 & -0.000 &  10.204 & 11.22 &                  &      O9-B \\
00227 & 266.41742 & -29.009571 & -1.548 &  -6.338 & 11.24 &                  &         ? \\
00366 & 266.41705 & -29.010412 & -0.570 &  -9.366 & 11.43 &                  &           \\
00445 & 266.41647 & -29.005707 &  0.981 &   7.574 & 11.58 &                  &      O9-B \\
00372 & 266.41827 & -29.008778 & -3.832 &  -3.481 & 11.60 &                  &           \\
00483 & 266.42029 & -29.005968 & -9.237 &   6.633 & 11.70 &          IRS 5SE &        B3 \\
00516 & 266.41754 & -29.006582 & -1.879 &   4.422 & 11.71 &                  &      B0-3 \\
00567 & 266.41855 & -29.006989 & -4.574 &   2.959 & 11.96 &                  &           \\
00507 & 266.41632 & -29.008602 &  1.386 &  -2.850 & 11.98 &                  &  O8.5-9.5 \\
00610 & 266.41849 & -29.009783 & -4.399 &  -7.100 & 12.01 &                  &           \\
00722 & 266.41489 & -29.010849 &  5.209 & -10.938 & 12.10 &                  &           \\
00757 & 266.41400 & -29.008787 &  7.583 &  -3.516 & 12.12 &                  &           \\
00508 & 266.41867 & -29.007784 & -4.897 &   0.096 & 12.13 &                  & O9.5-B2II \\
00617 & 266.41415 & -29.009539 &  7.171 &  -6.221 & 12.14 &                  &           \\
00721 & 266.41904 & -29.006376 & -5.884 &   5.164 & 12.22 &                  &           \\
00725 & 266.41602 & -29.008396 &  2.202 &  -2.108 & 12.31 &                  &     O9-B0 \\
00728 & 266.41425 & -29.008633 &  6.932 &  -2.959 & 12.31 &                  &           \\
00785 & 266.41705 & -29.008959 & -0.571 &  -4.134 & 12.34 &                  &      B0-1 \\
00707 & 266.41733 & -29.008621 & -1.305 &  -2.918 & 12.41 &                  &      B0-3 \\
00936 & 266.41479 & -29.009893 &  5.458 &  -7.498 & 12.42 &                  &           \\
00838 & 266.41455 & -29.006830 &  6.126 &   3.529 & 12.43 &                  &           \\
00853 & 266.41730 & -29.010958 & -1.221 & -11.330 & 12.45 &                  &           \\
\hline
\end{tabular}
\tablefoot{
The value for $K_{S}$ magnitudes are taken from \cite{schodel2010accurate}, while all other values originate from \cite{feldmeier2015kmos}. 
}
\end{table*}

\FloatBarrier

\section{SEDs of RC stars along FoV}\label{SED_fov}

\begin{figure*}[!ht]

\centering
\includegraphics[width=\textwidth]{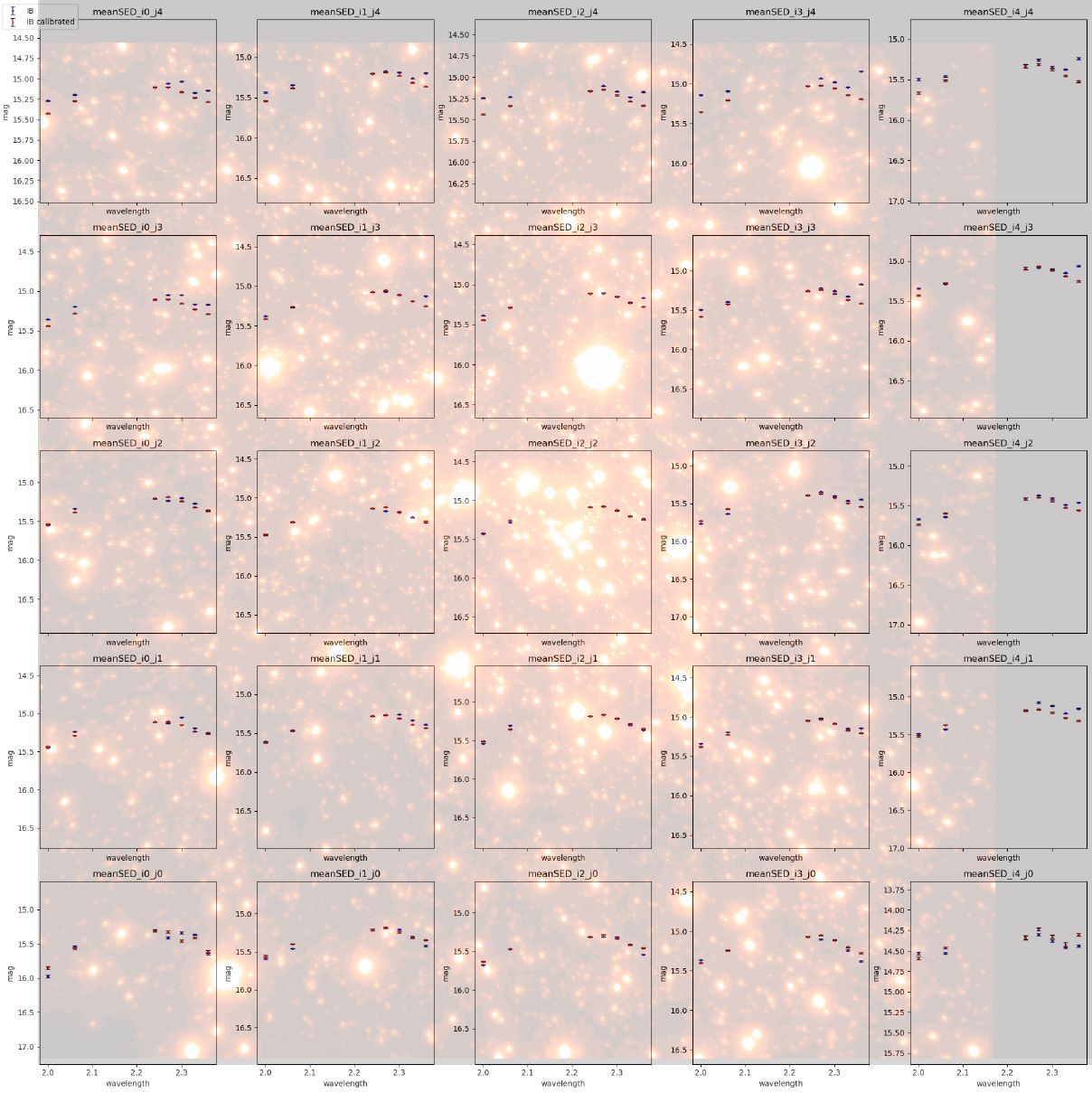}
\caption{Comparison between the mean SED of RC stars in the different sub-fields using the IB data after the basic calibration (blue points) and after applying the local calibration (red points). The SEDs are shifted in magnitude to achieve overlap. The effect of the local calibration is more important in the outer regions. The size of each sub-field is around $8.6''$ x $8.6''$. The panels corresponding to the different sub-fields overlap with the specific sub-field in the background image.}
  \label{Fig:sed_rc_fov}
\end{figure*}

In this section, we show a comparison between the mean SED of RC stars in the different sub-fields using the IB data after the basic calibration and after applying the local calibration along the FoV. In Fig.\,\ref{Fig:sed_rc_fov}, the influence of the local calibration is clearly illustrated, particularly in the outer regions. The blue points represent the SED after the basic calibration, while the red points depict the SED after applying the local calibration. The former is shifted in magnitude to achieve overlap.

\FloatBarrier

\section{Analysis of CBD fits}\label{CBD_fits_analysis}
\begin{figure} [ht!]
\includegraphics[width=\columnwidth]{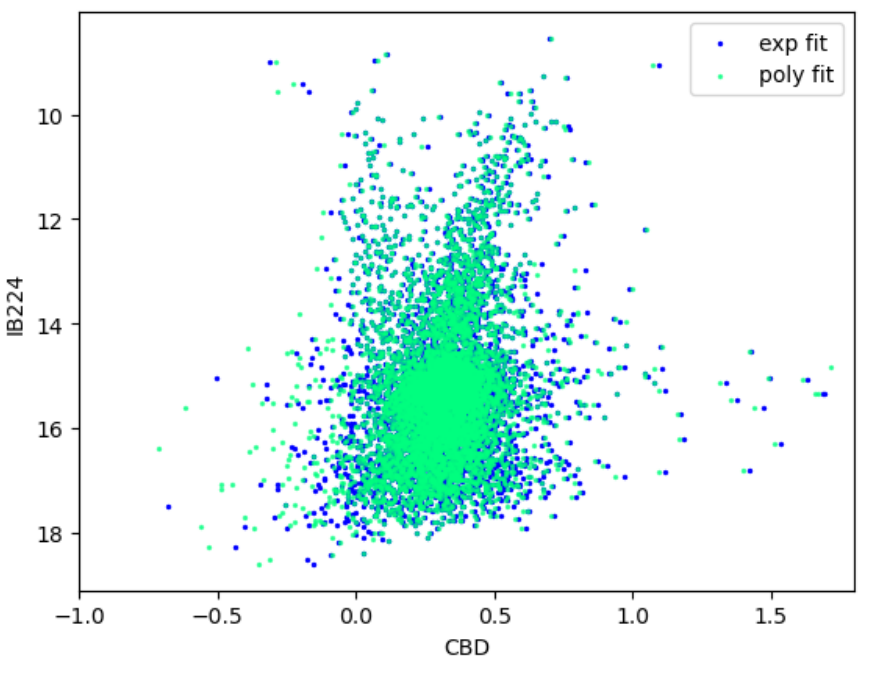}
\caption{Comparison of CBD diagrams obtained using two fitting methods: third-order polynomial (green points) and exponential function (blue points). }
  \label{Fig:comp_diagram_CBD_fits}  
\end{figure}

As we previously discussed in Section\,\ref{sec:cbd}, our method for measuring the CO band depth involves fitting a straight line to the four data points for each star. Then, we explored two different approaches to fit the entire SED of the stars: applying a third-order polynomial and fitting an exponential function (see Equation\,\ref{eqn:exp_fit}). Figure\,\ref{Fig:comp_diagram_CBD_fits} presents a comparison between the CBD diagrams obtained through these two methods. The distribution of the CBD values is very similar in both cases, except for some points on the left side of the diagram, where CBD values are approximately smaller than -0.1.

\begin{figure} [ht!]
\centering
\includegraphics[width=\columnwidth]{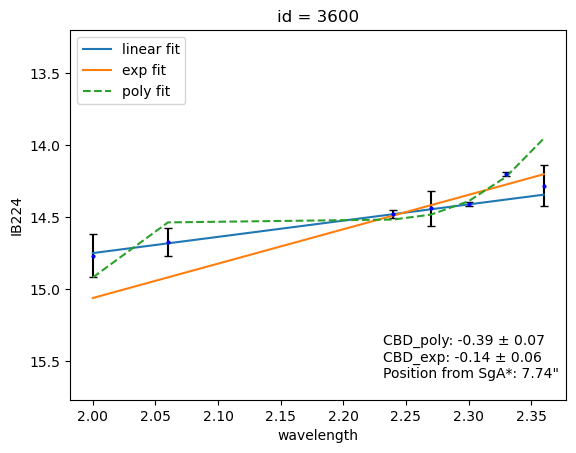}
\includegraphics[width=\columnwidth]{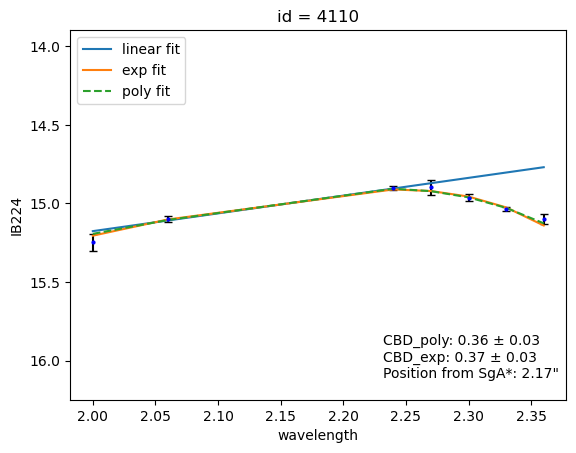} 
\caption{SEDs of two stars. The blue line denotes the linear fit to the first data points, the orange line represents the exponential fit, and the green line the polynomial fit to all seven points, respectively. The values of the CBD are annotated in the panels for the two fits.}
  \label{cbd_fits_comp}
\end{figure}

In Fig.\,\ref{cbd_fits_comp}, we compare the different fits for two stars. In situations where the SED lacks points with substantial errors or deviations from the overall shape, both fits appear highly similar (left panel). However, when the SED exhibits notable irregularities, the fits differ significantly, resulting in a discrepancy of more than 0.25 between the CBD values (right panel).

\FloatBarrier

\section{Exploring specific stars}\label{star_details}
\begin{figure} [ht!]
\includegraphics[width=\columnwidth]{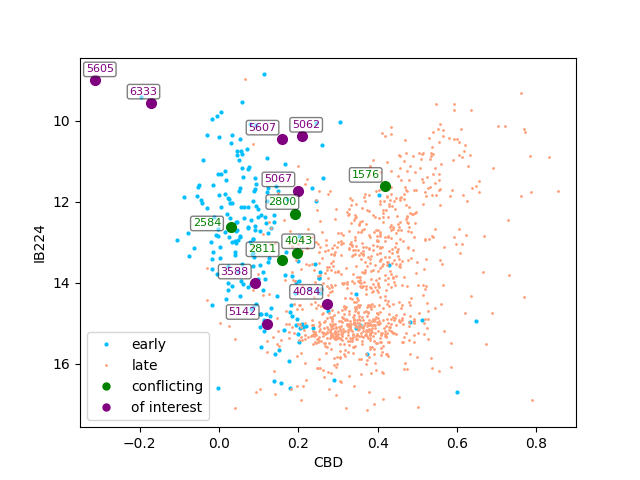}
\caption{CBD diagram highlighting stars of particular interest within our sample, including those with conflicting classifications in the literature and predominantly early-type candidates. Our Id labels are placed adjacent to the corresponding stars. Additionally, stars with spectroscopic literature classifications are included for comparison.}
  \label{Fig:special_stars}  
\end{figure}

In this section, we examine several stars of particular interest. Firstly, we analyzed five stars with conflicting spectral classifications and no apparent basis for preferring one over the other. We choose their types based on their positions in the CBDD (see Fig.\,\ref{Fig:special_stars}). The misclassification could be attributed to incorrect catalogue matching. We favour the late classification for only one star, our Id 1576, which corresponds to Id 553 in \cite{feldmeier2017kmos}, in contrast to the early classification assigned by \cite{yelda2014clockwise} as S13-3, based on the reuse of data from \cite{paumard2006two}. We note that this star is not categorized as early in \cite{fellenberg2022young}, the most recent MPE group work. The probabilities support this classification with a confidence level of 1.4\% or less for being classified as early. 

For the following four stars, we classified them as early. The first is Id 2584, also known as IRS7SE, which is consistently classified as early by three sources \citep{yelda2014clockwise, feldmeier2015kmos, fellenberg2022young}. The only exception is \cite{maness07old}, with their Id 37. The scatter in its SED is enhanced, however, enhanced scatter is not uncommon for stars with strong emission lines \citep[see][]{feldmeier2015kmos}.
The position in the CBDD (see Fig.\,\ref{Fig:special_stars}) suggests a clear tendency towards early classification, with probabilities ranging from 84\% to 95\%.
The next star, Id 2800, stands out as the only one among the four not classified as early (although there is no late classification) in the set of \cite{fellenberg2022young}. It is classified as early in \cite{stostad2015types} (N1-2-043) and as late in \cite{feldmeier2017kmos} (Id 448), although the latter dataset has a lower spatial resolution. Its classification as early ranges between 62\% and 76\%. The subsequent star, Id 2811, is classified as early in \cite{fellenberg2022young} (Id 27) and as late in \cite{feldmeier2017kmos} (Id 2156), with the latter dataset having lower spatial resolution. Its classification as early ranges between 71\%. and 80\%. The final star, Id 3043 \citep[also known as E87 from][]{paumard2006two}, is classified as early in \cite{fellenberg2022young} and \cite{yelda2014clockwise}, but late in \cite{maness07old}. Its classification probability leans towards early, ranging between 32\% and 41\%, which may be attributed to its relatively faint (13.3) and distant (11$''$) characteristics.

As we approached the conclusion of our study, we became aware of the work by \cite{jia23substructure}. After review, we found that four out of the six `new' early-type stars we identified correspond to stars in the \cite{fellenberg2022young} catalogue and, to some extent, in the \cite{yelda2014clockwise} catalogue. Specifically, the stars S11-246 and S5-106 are spectroscopic unclassified in our combined catalogue (designated as Our Id 3588 and 5142). Our early probabilities range from 76\% to 94\%, depending on the star and method used, thus affirming the validity of our classification methods.

Afterwards, we examined stars with only orbital data, provided by Stefan Gillessen (private communication). As anticipated, the faint central star S175 lacks a magnitude match due to high crowding. S145 corresponds to our Id 4084 and exhibits early probabilities ranging between 6\% and 38\%, consistent with its late spectral classification.

Subsequently, we examined bright stars without unusual extinction values. Among them, four stars brighter than magnitude 12 are more likely to be early according to Bayesian classification (IDs 5067, 5605, 5607, 6333). However, the certainty is not high, ranging from 61\% to 78\%, and there are significant differences in probabilities depending on the method, ranging from 23\% to 99\%, unlike most stars between magnitudes 12 and 14.5. This discrepancy is attributed to considerable scatter in the SED. It is possible that all except 5067 are located in the corners with only two exposures in some bands, and variability may also be a factor. While 5067 appears more promising, it still resides in the borderline region and could potentially be a metal-poor late-type star. Overall, we consider it unlikely that any of them are truly early, as bright early stars would likely have been discovered already.

Finally, we examined stars with unusual extinction values brighter than magnitude 12. All of them appear red, indicating they are either background objects or more likely intrinsically red. Except Id 5062 (IRS2L), all have impact classifications from \cite{feldmeier2015kmos} or \cite{feldmeier2017kmos}. Since our imaging offers somewhat higher spatial resolution than the best available spectroscopy,
it may assist in distinguishing the star's spectrum from its surrounding gas and dust. However, the practical implications are complex. For example, the SED of 5062 exhibits some signs of CO, albeit much weaker than usual, which could indicate dust-diluted CO. Nevertheless, red stars classified as early also exhibit a very similar SED. Additionally, the probability of early classification varies between 35\% and 98\% depending on the method used. In conclusion, higher spatial resolution spectroscopy may be necessary to penetrate the envelope and provide more conclusive insights.

\FloatBarrier

\section{Classification data}\label{classfication_data}
 
We provide the following columns: Id, coordinates, the corrected IB magnitudes and their errors, CBD and its error after all corrections, used spectroscopic classification, extinction flag, and a flag indicating if a star is located in a corner of the image, where data completeness may be lower. Additionally, we present probabilities for XGBoost and MLP, as well as likelihood, prior, and posterior for Bayesian analysis. Metallicity and temperature predictions are provided for stars within the validity of their prediction.

\section{Bayesian fits}\label{Bayesian_fit_details}
\begin{figure} [ht!]
\centering
\includegraphics[width=\columnwidth]{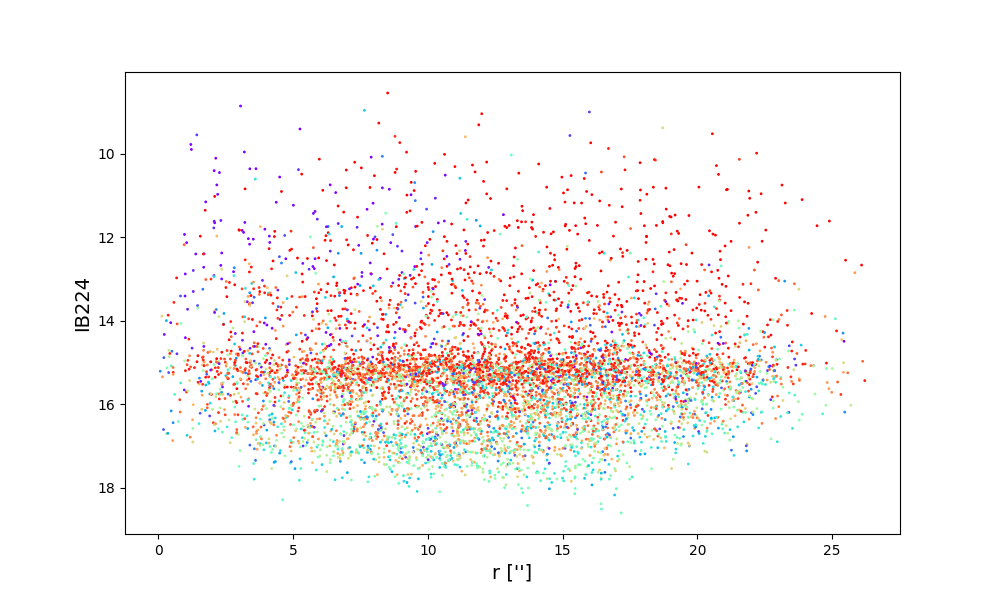}
\includegraphics[width=\columnwidth]{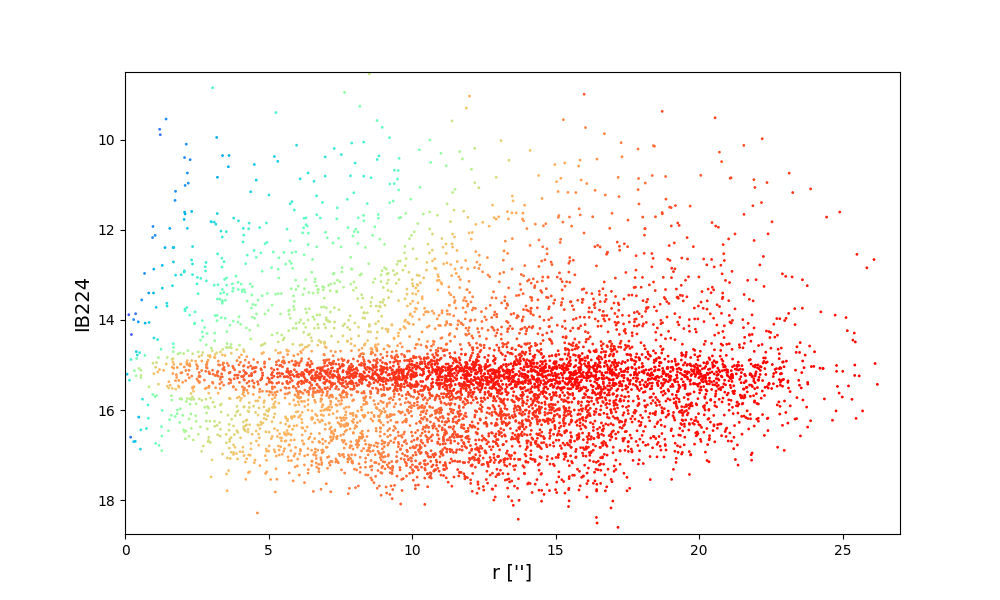}
\includegraphics[width=\columnwidth]{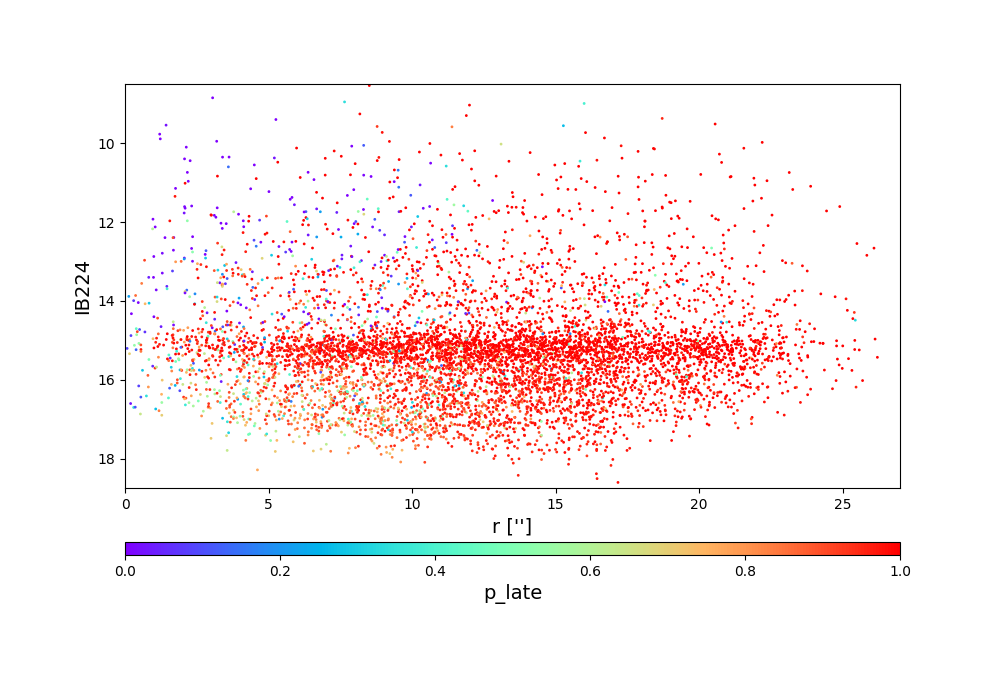} 
\caption{Likelihood, prior and posterior of our Bayesian model. In the absence of the prior (top panel), our model is intentionally designed to be independent of both radius and direct magnitudes. However, this changes when the prior (middle panel) is introduced (bottom panel). Then the radial trend becomes visible, and the RC stars are classified as late-type. All plots use the same colour scale for p\_late.}
  \label{Fig:mag_r_split}
\end{figure}

\begin{figure} [ht!]
\centering
\includegraphics[width=\columnwidth]{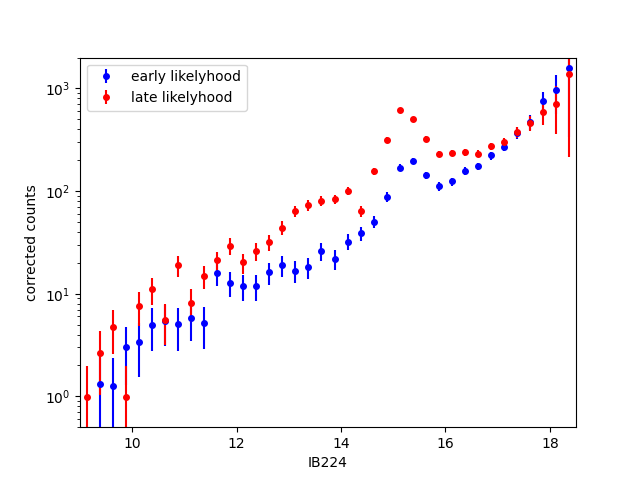}
\caption{KLF of the likelihood (in absence of the prior). The counts are completeness corrected for photometric incompleteness. There is a clear clump visible also for the early stars, although young stars do not show this clump. The reason is that in absence of a prior too many in reality late stars are classified early.}
  \label{Fig:klf_flat_prior}
\end{figure}

In this section, we present additional figures for the Bayesian analysis. First, we show the likelihood, prior, and posterior in magnitude-radial space in Fig.~\ref{Fig:mag_r_split}. We show the KLF of the likelihood in Figure~\ref{Fig:klf_flat_prior}.
 
\begin{figure*}
\centering
\includegraphics[width=\textwidth]{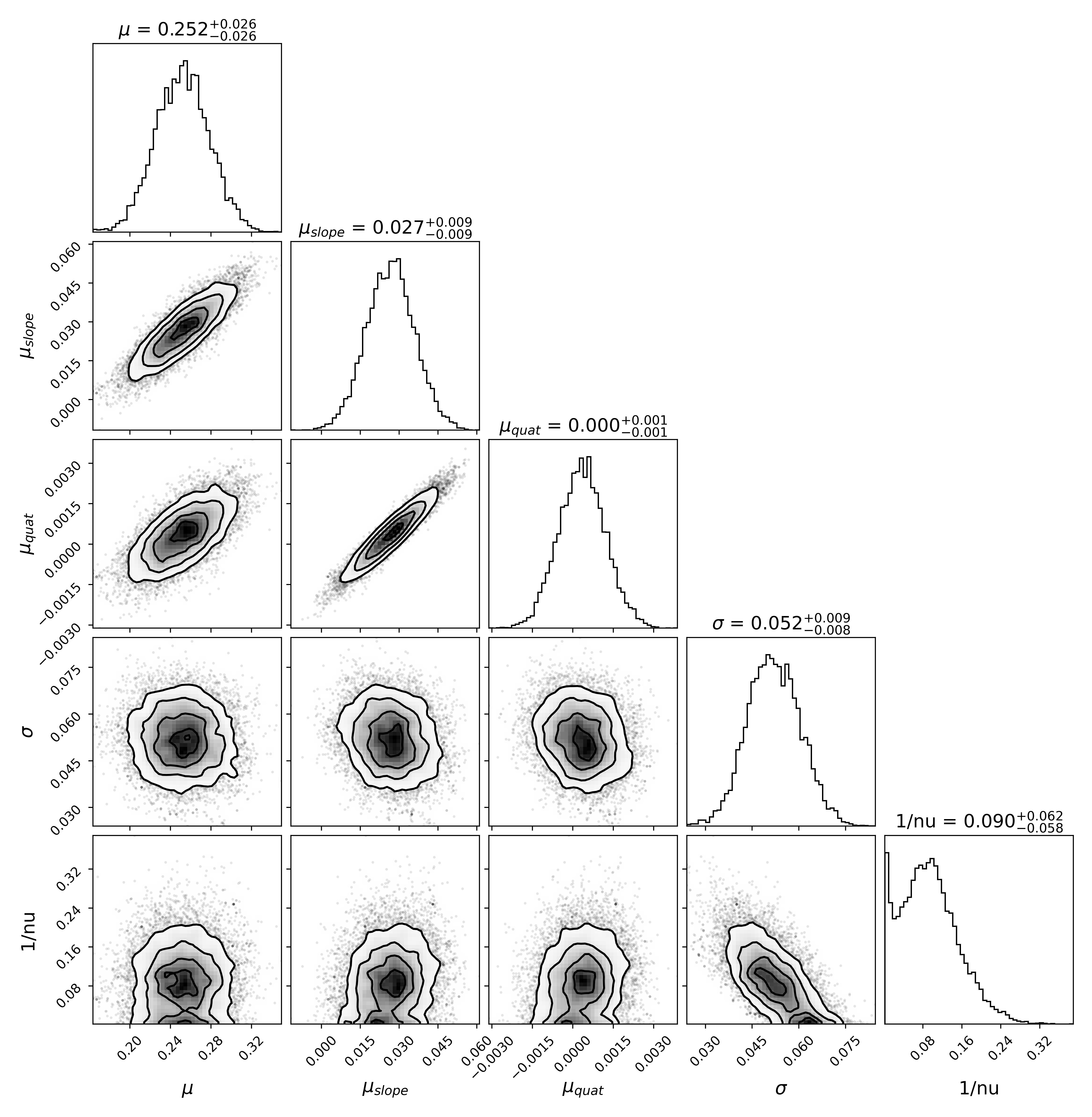}
\caption{Corner plot displaying the fit for early-type stars which employs a Student-t distribution.  In this context, $\nu$ parameterizes the deviation from a normal distribution and is equivalent to a normal distribution when $\nu=\infty$. By visualizing $1/\nu$, the transition to a normal distribution becomes more apparent and requires less space. It is important to underscore that the scale parameter $\sigma$ shares the same interpretation as the variance parameter in a normal distribution only when $\nu=\infty$.
The mean intercept $\mu$ is here defined at 19.5 mag and has a strong prior. This prevents the mean CBD of early stars from being larger than that of late stars, which is highly implausible.}
  \label{Fig:corner_early_student}  
\end{figure*}

\begin{figure*}
\centering
\includegraphics[width=\textwidth]{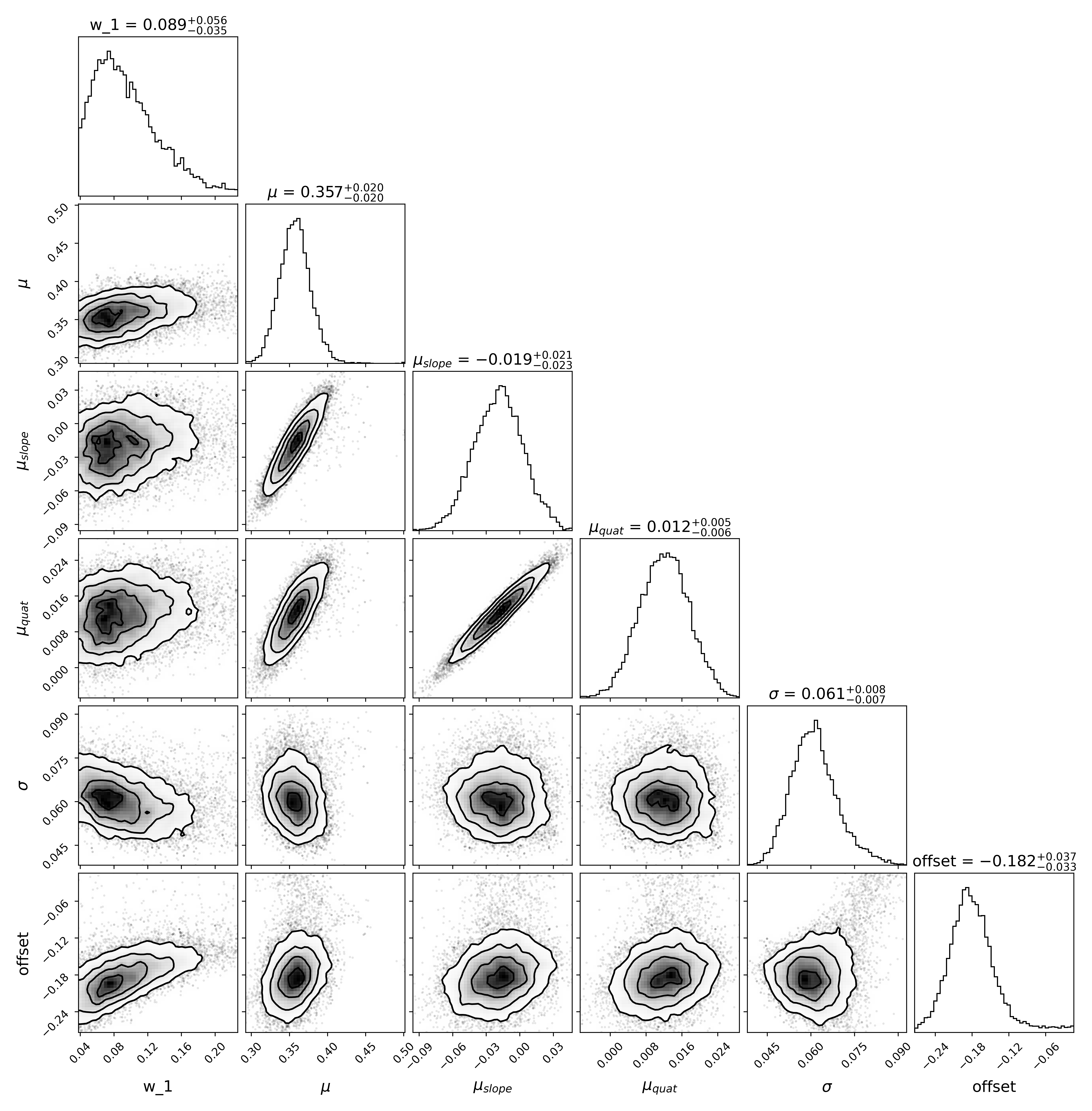}
\caption{Corner plot depicting our Bayesian fit for late-type stars brighter than 14.0. It employs 2 (offset) normal distribution fit for the late-type stars. A prior is used to ensure that component 1 is the minor one.}
  \label{Fig:corner_late_2student-t_b14}  
\end{figure*}

\begin{figure*}
\centering
\includegraphics[width=\textwidth]{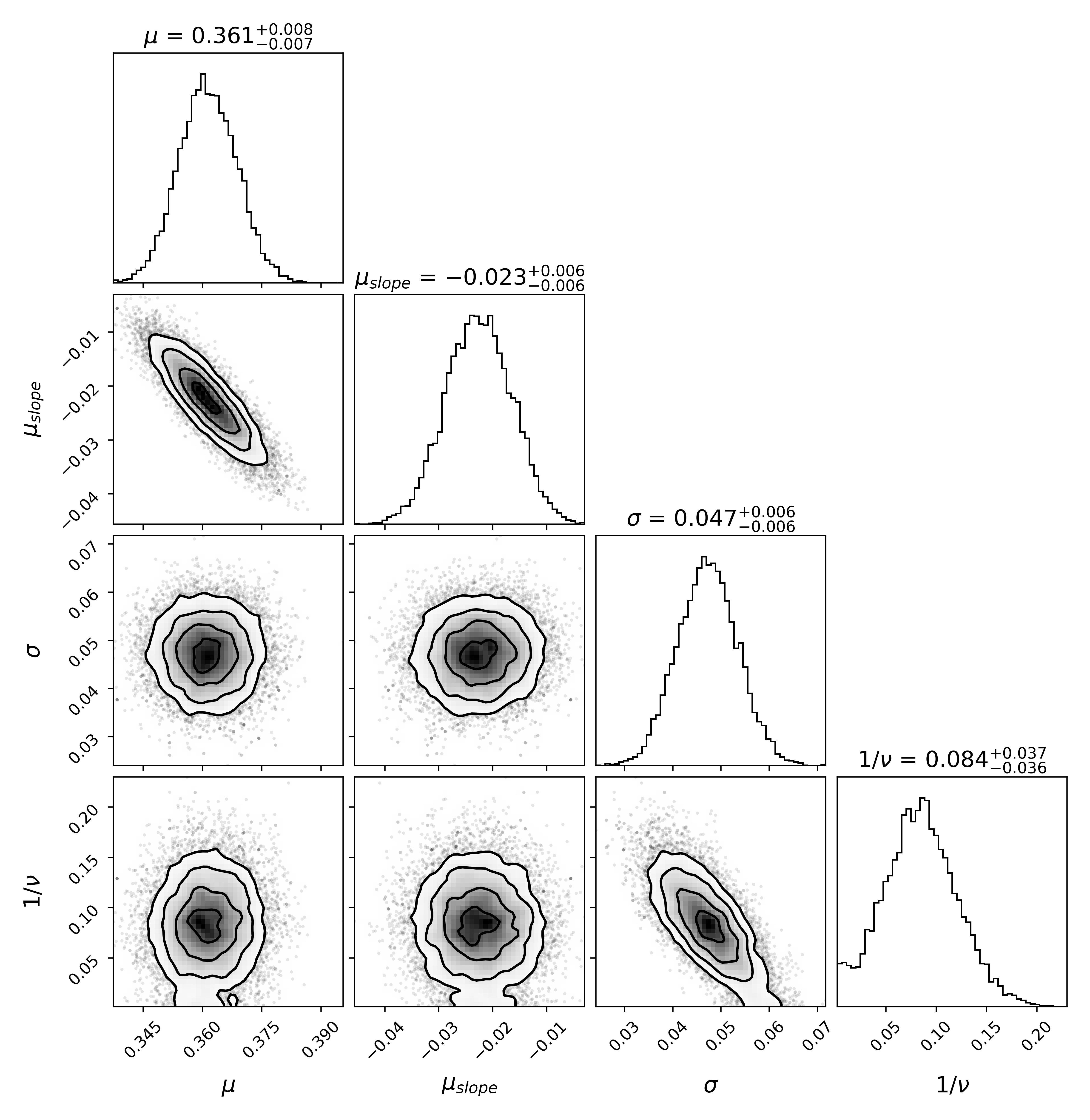}
\caption{Corner plot depicting our Bayesian fit for late-type stars fainter than 14.0 magnitudes.
It employs a Student-t distribution fit. In contrast to the other fits there is no quadratic parameter for the mean because it is not needed.}
  \label{Fig:corner_late_1student-t_f14}  
\end{figure*}

Then we present the corner plots for the Bayesian fits we employed. That is Student-t fit for early-type (Fig\,\ref{Fig:corner_early_student}) and for late-type brighter than 14 magnitudes the two components model of two normal distributions in Fig.~\ref{Fig:corner_late_2student-t_b14} and the 1 Student-t model for the late-type stars fainter than 14 magnitudes in Fig.~\ref{Fig:corner_late_1student-t_f14}.

\FloatBarrier
\clearpage 
\section{Multi-layer perceptron results}\label{MLP_results}

\begin{figure}[!ht]
\centering
\includegraphics[width=\columnwidth]{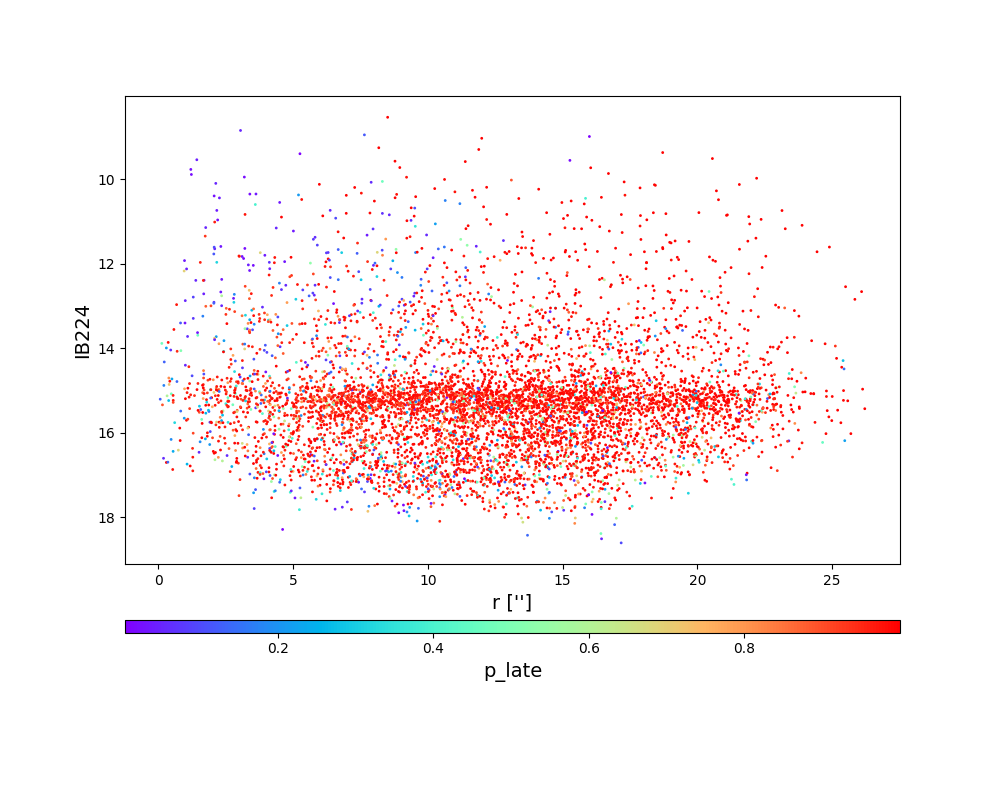} 
\caption{Probabilities of the stars obtained by the MLP method in CBD-magnitude-radius space, where the radius is the distance from Sgr~A*.}
  \label{Fig:nn_plots}
\end{figure}

 Figure ~\ref{Fig:nn_plots} shows the probabilities of the stars obtained by the MLP method in CBD-magnitude-radius space, where the radius is the distance from Sgr~A*.

\FloatBarrier

\section{Gradient boosted trees results}\label{GBT_results}

\begin{figure}[ht!]
\centering
\includegraphics[width=\columnwidth]{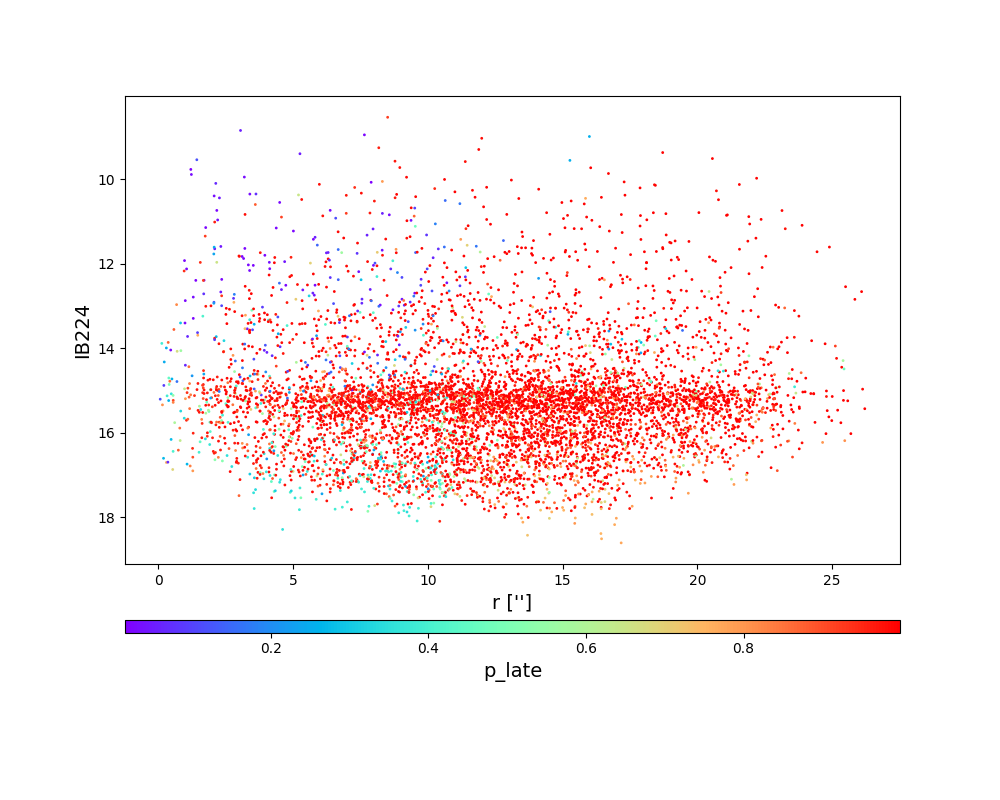} 
\caption{Probabilities obtained by the random forest in CBD-magnitude-radius space, the three most important features.}
  \label{Fig:rf_plots}
\end{figure}

Figure ~\ref{Fig:rf_plots} shows the probabilities obtained by the random forest in CBD-magnitude-radius space.

\FloatBarrier
\clearpage 
\section{SEDs of early-type candidates}\label{SED_early}

\begin{figure} [ht!]
\centering
\includegraphics[width=\columnwidth]{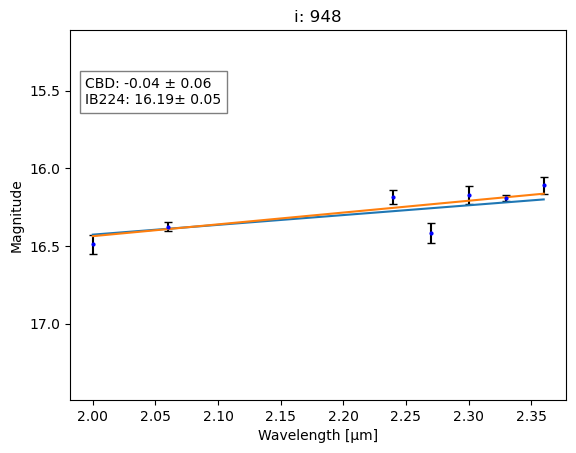}
\includegraphics[width=\columnwidth]{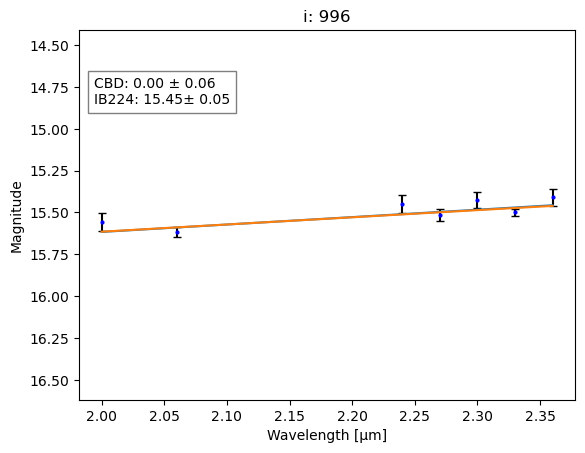}
\includegraphics[width=\columnwidth]{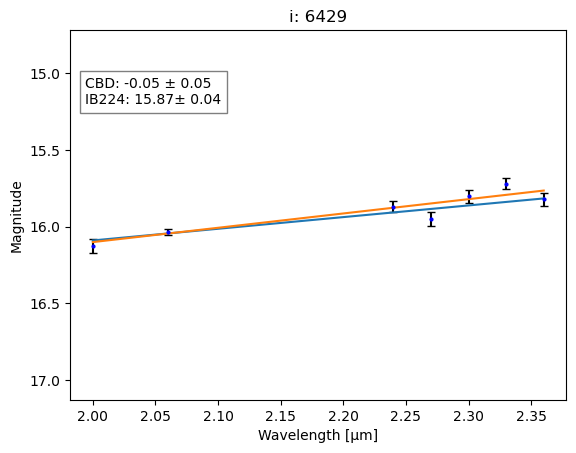}
    \caption{SEDs for early-type candidates detected only with MLP method, located in the upper left corner (upper panel), lower left corner (middle panel), and upper right corner (bottom panel).}
  \label{fig:sed_young_cand_mlp}
\end{figure}

Due to discrepancies in the number of early-type candidates among the three methods, particularly in the outer regions where the MLP method identifies a higher number of candidates compared to the other two methods, this appendix delves into the SEDs of stars detected exclusively with MLP. The goal is to assess the reliability of the results. Figure~\ref{fig:sed_young_cand_mlp} displays the SEDs of three early candidates located in the corners of the image (see Fig.~\ref{Fig:map_candidates}). The CBD values, as well as the shapes of their SEDs, suggest characteristics typical of early-type stars. Additionally, in the same regions, late-type candidates exhibit distinct SED features, including a noticeable dip corresponding to the CO band in wavelengths larger than 2.27. This observation leads us to the conclusion that some of the stars detected only by the MLP method may indeed correspond to genuine young stars.

\FloatBarrier

\section{On metallicity}
\label{app:metallicity}
\begin{figure} [ht!]
\begin{center}
\includegraphics[width=\columnwidth]{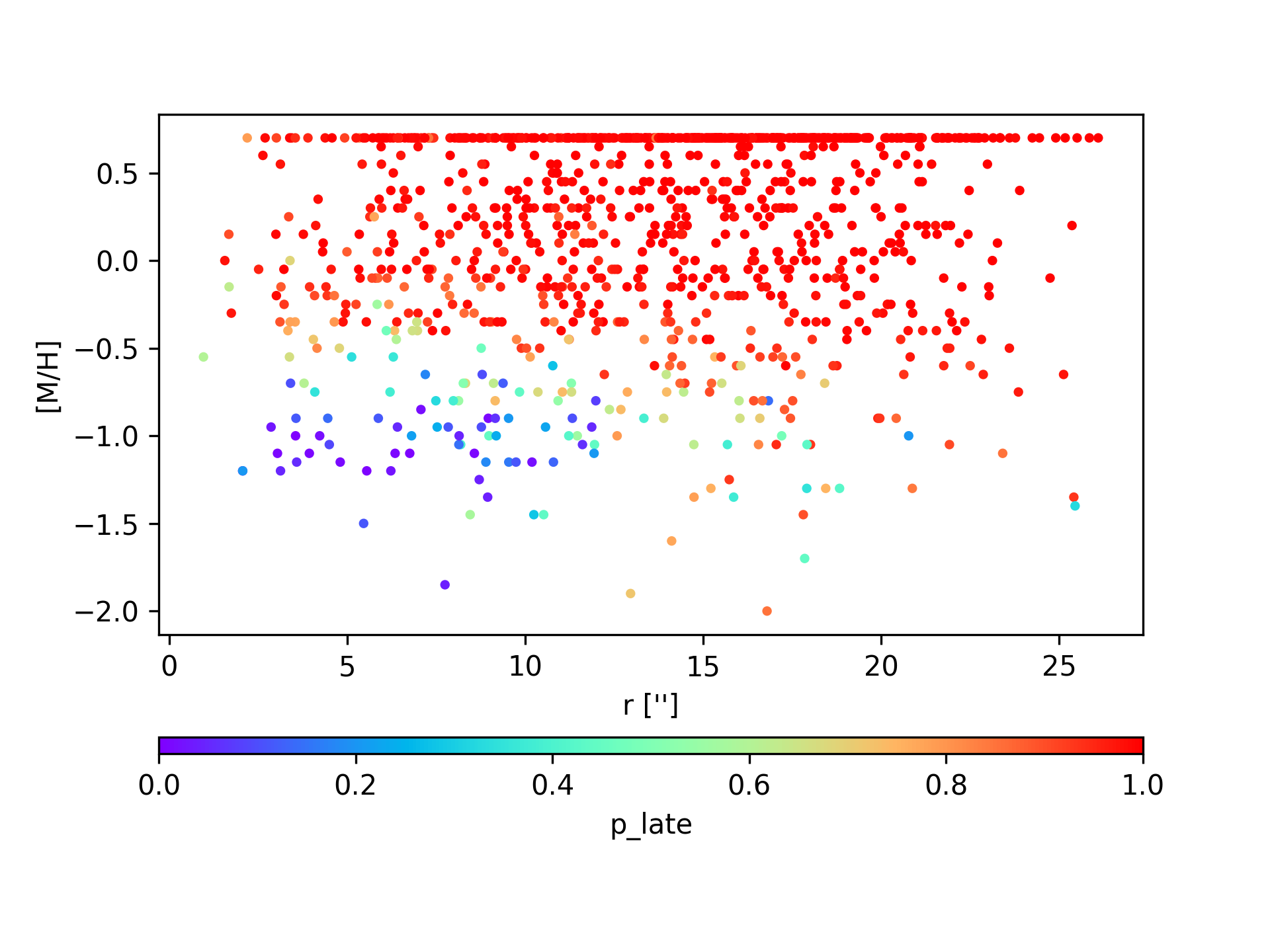}

\end{center}
    \caption{Predicted metallicity. Excluded are stars which are spectroscopically early and stars outside the magnitude range where the prediction works}
  \label{fig:met_radial}
\end{figure}

We show in Fig.~\ref{fig:met_radial} the interaction of predicted metallicity and late probability. Both need to be modelled at once for reliable results. 

\end{appendix}
\end{document}